\documentclass[final,times,twocolumn]{elsarticle}

\usepackage[english]{babel}
\usepackage{comment}
\usepackage{amssymb}

\usepackage{listings}
\usepackage[x11names]{xcolor}

\definecolor{backcolour}{rgb}{0.95,0.95,0.92}
\definecolor{codegreen}{rgb}{0,0.6,0}

\definecolor{codegray}{rgb}{0.5,0.5,0.5}
\definecolor{commentcolour}{rgb}{0.43,0.63,0.65}
\definecolor{shadecolor}{rgb}{0.93, 0.93, 0.93}
\definecolor{darkgreen}{rgb}{0.0, 0.5, 0.0}
\definecolor{darkred}{rgb}{0.8, 0.0, 0.0}

\lstdefinestyle{Python}{language=Python,    
    backgroundcolor=\color{shadecolor},
    commentstyle=\color{commentcolour},
    keywordstyle=\color{darkgreen},
    numberstyle=\tiny\color{codegray},
    stringstyle=\color{darkred},
    basicstyle=\ttfamily\footnotesize,
    breakatwhitespace=false,
    breaklines=true,
    captionpos=b,
    keepspaces=true,
    numbers=left,
    numbersep=3pt,
    showspaces=false,
    showstringspaces=false,
    showtabs=false,
    tabsize=2
}

\definecolor{cool}{RGB}{32,101,171}
\definecolor{hot}{RGB}{177,23,42}
\usepackage[pdftex,colorlinks=true, pdfstartview=FitV, linkcolor= hot, citecolor= hot, urlcolor= cool, hyperindex=true,hyperfigures=true]{hyperref}
\usepackage{amsmath}
\usepackage{wasysym}

\usepackage{lineno}
\usepackage[utf8]{inputenc}

\usepackage{float}
\restylefloat{table}

\usepackage[normalem]{ulem}

\definecolor{color_comment}{rgb}{0.8, 0.3, 0.3}

\definecolor{color_out}{rgb}{0.7, 0.7, 0.7}

\definecolor{color_new}{rgb}{0.3, 0.8, 0.3}

\biboptions{sort&compress}

\journal{Computer Physics Communications}

\begin{document}

\setlength{\parindent}{0pt}

\begin{frontmatter}



\title{QuOCS: The Quantum Optimal Control Suite}


\author[ulm,padova,infn,qruise]{Marco Rossignolo\corref{cor1}\fnref{fn1}}
\ead{marco@qruise.com}
\author[juelich,koln]{Thomas Reisser\fnref{fn1}}
\author[ulm,qruise,nvis]{Alastair Marshall}
\author[padova,infn,koln]{Phila Rembold}
\author[padova,infn,ulm2]{Alice Pagano}
\author[ulm]{Philipp J. Vetter}
\author[ulm]{Ressa S. Said}
\author[juelich]{Matthias M. Müller}
\author[juelich,koln]{Felix Motzoi}
\author[juelich,koln]{Tommaso Calarco}
\author[ulm]{Fedor Jelezko}
\author[padova,infn]{Simone Montangero}

\address[ulm]{Institute for Quantum Optics, Ulm University, Albert-Einstein-Allee 11, 89081 Ulm, Germany}
\address[padova]{Dipartimento di Fisica e Astronomia "G. Galilei" \& Padua Quantum Technologies
Research Center, Università degli Studi di Padova, Italy I-35131, Padova, Italy}
\address[infn]{INFN, Sezione di Padova, via Marzolo 8, I-35131, Padova, Italy}
\address[qruise]{Qruise GmbH, Saarbrücken, D-66113, Germany}
\address[juelich]{Peter Grünberg Institute -- Quantum Control (PGI-8), Forschungszentrum J\"ulich GmbH, D-52425 Germany}
\address[koln]{Institute for Theoretical Physics, University of Cologne, D-50937 Germany}
\address[nvis]{NVision Imaging Technologies, Albert-Einstein-Allee 11, 89081 Ulm, Germany}
\address[ulm2]{Institute for Complex Quantum Systems \& Center for Integrated Quantum Science and Technology, Ulm University, Albert-Einstein-Allee 11, 89081 Ulm, Germany}

\cortext[cor1]{Corresponding author}

\fntext[fn1]{These authors contributed equally to this work.}

\begin{abstract}
Quantum optimal control includes a family of pulse-shaping algorithms that aim to unlock the full potential of a variety of quantum technologies. 
The Quantum Optimal Control Suite (QuOCS) unites experimental focus and model-based approaches in a unified framework. 
Easy usage and installation presented here and the availability of various combinable optimization strategies is designed to improve the performance of many quantum technology platforms, such as color defects in diamond, superconducting qubits, atom- or ion-based quantum computers. It can also be applied to the study of more general phenomena in physics.
In this paper, we describe the software and the toolbox of gradient-free and gradient-based algorithms. We then show how the user can connect it to their experiment. 
In addition, we provide illustrative examples where our optimization suite solves typical quantum optimal control problems, in both open- and closed-loop settings.
Integration into existing experimental control software is already provided for the experiment control software Qudi [J. M. Binder et al., SoftwareX, 6, 85-90, (2017)], and further extensions are investigated and highly encouraged.
\\
QuOCS is available from \href{https://github.com/Quantum-OCS/QuOCS}{GitHub}, under Apache License 2.0, and can be found on the \href{https://pypi.org/project/quocs-lib}{PyPI} repository.

\end{abstract}



\end{frontmatter}


\vspace{2em}


{\bf Program summary}

\begin{small}
\noindent
{\em Program Title:} QuOCS - Quantum Optimal Control Suite\\
{\em CPC Library link to program files:} (to be added by Technical Editor) \\
{\em Developer's repository link:} \\
    \href{https://github.com/Quantum-OCS/QuOCS}{https://github.com/Quantum-OCS/QuOCS} \\
{\em Code Ocean capsule:} (to be added by Technical Editor)\\
{\em Licensing provisions:} Apache-2.0 \\
{\em Programming language:} Python\\
{\em External routines:} NumPy~\cite{NumPy}, SciPy~\cite{NumPy}, JAX~\cite{JAX}\\
{\em Nature of problem:}\\
    Quantum systems are typically controlled by time-dependent electromagnetic fields to perform a certain set of quantum operations. Those operations may in turn provide building blocks for various quantum information processing tasks such as quantum computation, communication, simulation, sensing, and metrology. Numerous control strategies exist to design and improve such operations [3]. While some strategies are constructed to target a rather specific problem with high efficiency, others are more general to solve a wide range of applications [4]. To access the different algorithms, one has to download different optimization suites with different input and output parameters which makes them hard to compare and harder to combine. To benefit from the variety of algorithms, we have devised a customizable and intuitive optimization suite that simultaneously provides access to some of the most popular quantum optimal control algorithms.\\
{\em Solution method:}\\
    We combine, in a unified framework, some of the frequently used optimal control algorithms which are the dressed Chopped Random Basis method (dCRAB) [5], and Gradient Ascent Pulse Engineering (GRAPE) [6], with an extension to make use of Automatic Differentiation (AD) [7]. The software is able to connect to both models of quantum dynamics in simulations and real-time quantum experiments to perform open- and closed-loop optimization, respectively. With minimal knowledge of optimal control theory, the user can manage to run optimizations of quantum processes using a variety of additional features such as stopping criteria and drift compensation. Logging and data management of the optimization progress and results are also handled by the suite. Its modular structure allows for extensions that accommodate customized algorithms and can be implemented by the user straightforwardly. \\
{\em Additional comments including unusual features:}\\
    The connection to the experiments is performed by an extension that enables a direct integration to a laboratory control software Qudi [8]. \\

\end{small}



\section{Introduction}
\label{sec:Intro}






Quantum optimal control (QOC) is a technique for shaping pulses that can enable and improve the performance of quantum technology and contribute to studying phenomena in quantum physics in general~\cite{Brif_2010, Glaser2015, Koch_2016, Koch_2022, Mueller2022, Rembold2020}. 
Besides a clever way of combining search algorithms with modeled or measured information about a quantum system, it includes analytical methods such as Pontryagin's maximum principle~\cite{Gamkrelidze_1999, Pesch_2012}, geometric approaches~\cite{Geometric_Control_2002, Clark_2021}, shortcuts to adiabaticity~\cite{STA_1997, STA_2003, STA_2010, STA_2013, STA_2019, DRAG_2009} and dynamical decoupling~\cite{DD_1950, DD_1954, DD_1999, DD_2003, DD_2005, DD_2009, DD_2010, Degen_2017}.
QOC can be used to develop control sequences for quantum systems~\cite{muller2018noise}. This can include guiding the system from an initial state to some desired target state~\cite{Petruhanov_2022}, producing robust quantum gates, or entanglement~\cite{Preti_2022}. To do so, the algorithm either makes use of a simulated model or adaptive learning from the experiment. If optimal control is used in combination with a simulation and theoretical model, we refer to it as open-loop quantum optimal control. If, however, the optimization is performed using information obtained from experimental measurements of the controls provided by QuOCS, we speak of a closed-loop optimization. Closed-loop QOC is primarily applicable to experiments that use variable electromagnetic control fields and are capable of performing fast state preparation and read-out compared to the lifetime and coherence time of the system. When using QOC for closed-loop optimization, regardless of the origin of the feedback, the algorithm iteratively advances the controls until the system reaches the desired target in a reasonable computational and experimental time.

QOC has its roots in the nuclear magnetic resonance community of the 1970s, where, for example, shaped radio frequency pulses were designed to display certain spectral properties~\cite{tomlinsonFourierSynthesizedExcitation1973}. Since then, many methods have been developed specifically catering to quantum problems: 
The Krotov method~\cite{Tannor1992, goerzKrotovPythonImplementation2019, reichMonotonicallyConvergentOptimization2012} is a gradient-based approach that can under certain assumptions guarantee monotonic convergence.
The field has steadily grown, and with the introduction of the gradient ascent pulse engineering (GRAPE) method in 2005, optimal control became more accessible~\cite{khaneja2005optimal, Machnes2011}.
In 2011, the chopped random basis (CRAB) algorithm~\cite{Doria_2011, canevaChoppedRandombasisQuantum2011a} was introduced and refined in 2015 with the introduction of dressed CRAB (dCRAB)~\cite{rachDressingChoppedrandombasisOptimization2015a}. In the open-source domain, many libraries have implemented algorithms similar to the ones we present here, e.g., the quantum toolkit in Python (QuTiP)~\cite{johansson2012qutip} includes both CRAB and GRAPE, the \texttt{Krotov.py} package~\cite{goerzKrotovPythonImplementation2019} contains a Python implementation of the Krotov method. QOpt offers an experiment-oriented qubit simulation and optimization control package~\cite{teskeQoptExperimentorientedQubit2022}, DYNAMO~\cite{Machnes2011} and Spinach~\cite{HOGBEN2011179} are MATLAB-based programs running GRAPE, and C$^3$ is an integrated open-source tool-set for Control, Calibration, and Characterization~\cite{Wittler_2021}. 
In particular, the main focus of C$^3$ is on superconducting qubit systems and relies on the description of the system via Hamiltonians. 
What sets the Quantum Optimal Control Suite (QuOCS) apart is that it unites model-based approaches and an (also model-free) experimental focus to provide a unified framework. The common interface and large selection of customizable options make it an ideal platform for application-focused yet efficient optimization.
At its core, QuOCS already includes a standard set of gradient-based and gradient-free methods. However, it is the whole environment QuOCS offers to make this software noteworthy: it allows the user to create and benchmark their own tailored QOC algorithms. Furthermore, it is designed to be easily connected to lab control setups (leveraging existing experimental control software such as Qudi~\cite{Binder2017}) and thus be used as a closed-loop optimizer.
QOC algorithms have already demonstrated their capabilities in many different physical platforms, including NV centers in diamond~\cite{scheuer2014precise, waldherr2014quantum, dolde2014high, unden2016quantum, Frank2017, schmitt2017submillihertz, poggiali2018optimal, muller2018noise, Oshnik_2022}, nuclear magnetic resonance (NMR)~\cite{Conolly_1986, Peirce_1988, McDonald_1991, khaneja2005optimal, Casanova_2018, Vetter_2021}, trapped ions~\cite{muller2015phonon, furst2014controlling, pichler2016noise, monz201114, walther2012controlling, casanova2011quantum, singer2010colloquium, zhang2018noon, leibfried2003quantum}, cold atoms~\cite{rosi2013fast, van2014interferometry, van2016optimal, brouzos2015quantum, sorensen2018quantum, heck2018remote, omran2019generation, Mastroserio2022}, and superconducting qubits~\cite{watts2015optimizing, goerz2015optimizing, hoeb2017amplification, Heeres2017, Theis_2018}. The optimization tasks range from high-performance sensing and state preparation to the creation of controlled unitary operations for quantum computing and quantum simulation.
QuOCS inherits ideas from RedCRAB (Remote dressed Chopped RAndom Basis)~\cite{canevaChoppedRandombasisQuantum2011a, rachDressingChoppedrandombasisOptimization2015a} on optimizing theoretical problems and experimental challenges~\cite{Heck2018, Angaroni_2019, macroscopic_hyperpol_2022} concentrating on a user-friendly and customizable interface. In contrast to RedCRAB, QuOCS is an open-source software with a modular structure and object-oriented nature providing a wider range of customizability to the user. QuOCS has already found its application in the optimization of a two-qubit gate with Rydberg atoms~\cite{Pagano2022} and in this work we demonstrate a closed-loop application with NV centers in diamond.\\

The paper is structured as follows. In Sec.~\ref{sec:software_description} we describe the software, the main toolbox and show how to connect to an experiment. Illustrative examples where our optimization suite solves typical quantum optimal control problems, both open- and closed-loop, are given in Sec.~\ref{sec:example_applications}. Finally, in Sec.~\ref{sec:conclusions_and_outlook} we give an outlook and a summary of this work.

\section{Software Description}
\label{sec:software_description}
 
 

Python has gathered a lot of attention in the last decade, especially within the scientific community. As an intuitive, easy-to-read, and flexible programming language, it has become the de facto standard for writing user-friendly code. Whenever its efficiency is not sufficient, a number of application programming interface (API) libraries can be used to connect it to other high-performance programming languages for computationally heavy calculations. To leverage this existing community, QuOCS is written in Python 3 using an object-oriented style and has a robust suite of tests to ensure the code is reliable. This chapter leads the reader through a QOC process, starting from the problem description, through the configuration of the optimizer and ending with the modules that make up the software and available algorithms.

To get started with QuOCS the user should identify the available time-dependent and static controls and their limits. Additionally, a rigorous definition of the specified target is required.
The next step is to set up the connection between the system (experiment or simulation) and QuOCS. At this point, the user has to choose the optimal control algorithm and hyperparameters before starting the iterative optimization. Ultimately, the algorithm converges to produce the improved final control sequence.

\subsection{Problem Description}
\label{sec:problem_description}
A QOC problem is generally described through the dynamics of a quantum system and an optimization target. The Hamiltonian of a quantum system, 
\begin{equation}
\label{eq:QOC_problem_hamiltonian}
H(t) = H_0(t) + \sum_j u_j(t) \cdot H_{c,j}(t) \; ,
\end{equation}
is typically defined by a system term $H_0$, also called the drift Hamiltonian, and the control Hamiltonian $H_{c,i}$, which represents the effect the control fields $u_i(t)$ have on the system. The system starts in an initial state $|\Psi_0\rangle$ or $\rho_0 = |\Psi_0\rangle \langle\Psi_0|$ and evolves according to the time-dependent Schrödinger equation
\begin{equation}
\label{eq:schroedinger_eqn}
\frac{\partial}{\partial t} \, |\Psi(t)\rangle = - \frac{i}{\hbar} \, H(t) \, |\Psi(t)\rangle
\end{equation}
or more generally the Liouville-von Neumann equation.
\begin{equation}
\label{eq:liouville_eqn}
\frac{\partial}{\partial t} \, \rho(t) = - \frac{i}{\hbar} \, \left[H(t), \rho(t) \right] \; .
\end{equation}
The latter can be extended to include Lindblad terms to describe decoherence and dissipation~\cite{Marquardt_2008}. In the case of a closed quantum system, the solution to equations (\ref{eq:schroedinger_eqn}) and (\ref{eq:liouville_eqn}) is given by
\begin{equation}
\label{eq:sol_schroedinger}
|\Psi(t)\rangle = U(t) \, |\Psi_0\rangle
\end{equation}
with the unitary propagator,
\begin{equation}
\label{eq:propagator}
U(t) = \mathbb{T} \, \exp\left(-\int_0^t H(\tau) d\tau\right)
\end{equation}
where $\mathbb{T}$ is the Dyson time-ordering operator~\cite{Machnes2011}.
The preparation of a quantum state can typically be done by intrinsic properties of the given quantum system, but can also be the objective of a QOC problem. A comparison to the target is done via the Figure of Merit (FoM) which quantifies the distance to the target after the evolution or its preparation fidelity. Such a FoM can be the state overlap fidelity, $F_{\text{state}}~=~\left( \text{tr} \sqrt{\sqrt{\rho_T} \; \rho_{\text{aim}}\sqrt{\rho_T}} \right)^2$, or gate overlap fidelity, $F_{\text{gate}}~=~\frac{1}{N^2} \text{tr}\left( U_{\text{aim}}^{\dagger} \, U_T  \right)^2$. It can also be extended to contain further information such as penalty terms. Such term can, for example, help to reduce the population of forbidden or undesired (usually lossy) states. Alternatively, any term can be included that represents a parameter that should be maximized or minimized during the development of the control sequence. Other examples include the control power or high-frequency pulse components. In order to limit the power transmitted to the system by the control, a factor $k$ can be introduced such that the FoM takes the form $ \mathcal{F} = k \cdot (1 - F)  + (1 - k) \int |u(t)|^2 dt $ where $F$ is e.g. the gate fidelity. In that way it becomes clear that the FoM is not necessarily the same as the (in)fidelity, and the optimization can be adjusted to be in favor of increasing the fidelity or reducing the power. During the optimization, a FoM is calculated in each iteration of the search and associated with the tested control pulse. The pulses $u_j(t)$ (or control fields) are expanded in an arbitrary type of basis, such as, e.g., the Fourier basis or Chebyshev polynomials, and updated by variation of the basis coefficients (i.e., the optimization-parameters). Additionally, the gradient of the FoM can be calculated with respect to the basis elements of the pulse. 

As an example for the user, we have implemented a gradient-based optimization for a state-to-state transfer, where the user sets the initial and target state as well as the model of the quantum system by providing the drift and control Hamiltonian. In that case, QuOCS also works as a simulator. 
When performing gradient-free optimizations, the user only has to provide the custom FoM class, which is a child of the \texttt{AbstractFoM} class, that feeds back the fidelity or cost of a given set of pulses and parameters. The system dynamics are provided by a user-specific coded model. Future extensions to the software might include a set of efficient propagators, sophisticated solvers of the Schrödinger equation, or approximations for ensemble systems if the problem is described in the template structure provided by QuOCS.
Regardless of the specific way the FoM is calculated, the aim of the \texttt{AbstractFoM} class is to establish a connection to any kind of simulation code or experiment. In the simplest case, this can be done by calling another Python code script from within the FoM class. 
In principle, any arbitrary structure that takes the pulses, their associated times, and constant but variable parameters as an input, can work for a gradient-free black-box optimization with QuOCS. The only requirement is that a real number to be maximized or minimized is given as an output.
As a result, the user can execute code written in other programming languages such as Julia, C, C++, Matlab, or Mathematica. For connections to an experiment, methods of file communication, where the pulses, parameters, and FoM are written into text files and a watcher for timestamp changes is implemented, have been proven very useful in the development and experiences with RedCRAB. Furthermore, TCP/IP socket communication, where the information is sent between local ports on the same computer or remotely accessible ports between different machines, has been investigated and might become available in future updates to the software.

Once the FoM has been calculated for a given set of parameters describing the pulse, QuOCS uses an updating algorithm (such as the Nelder-Mead simplex algorithm~\cite{Nelder1965,gao2012implementing}, Powell's method \cite{Powell1964}, the covariance matrix adaption evolution strategy (CMA-ES)~\cite{Hansen2003}, or L-BFGS-B~\cite{Liu1989}) to find a new pulse. This cycle is repeated until the optimization converges.

\subsection{Configuration of the Optimizer}
\label{sec:Configuration}
After defining the problem to be solved via the \texttt{AbstractFoM} class, the settings for the solution method of the problem by QuOCS have to be specified by the user. This is done in a configuration file with JSON format. JSON provides a structured and human-readable way of defining data as a dictionary in a Pythonic style. We have split the information into sections to ease the navigation between different settings and the parameters provide intuitive configuration. Examples can be found in Sec.~\ref{sec:example_applications} and in the form of a set of \href{https://github.com/Quantum-OCS/QuOCS-jupyternotebooks}{Jupyter notebooks on GitHub}~\cite{jupyter_notebooks}. The configuration file is saved automatically together with the optimization results to allow the user to take note of the used parameters and reproduce the results. We will go through the settings in the order as shown in the examples. Each key is an entry in the optimization configuration dictionary and contains the information of a specific setting.
A name for the optimization runs, that is then also contained in the folder name for the results, can be specified by the key \colorbox{Snow2}{\texttt{optimization\_client\_name}}. The main sections for the optimization are \colorbox{Snow2}{\texttt{algorithm\_settings}}, \colorbox{Snow2}{\texttt{pulses}}, \colorbox{Snow2}{\texttt{parameters}} and \colorbox{Snow2}{\texttt{times}}. The \colorbox{Snow2}{\texttt{algorithm\_settings}} define the information about the type of QOC algorithm and its properties. Among them are, for example, the name of the algorithm (e.g. dCRAB, GRAPE, or AD (for automatic differentiation) (see Sec.~\ref{app:qoc_algorithms})) via the key \colorbox{Snow2}{\texttt{algorithm\_name}}, the number of super-iterations in the case of dCRAB (see \ref{app:dCRAB}) by \colorbox{Snow2}{\texttt{super\_iteration\_number}}, and if a minimization or maximization of the FoM should be performed (\colorbox{Snow2}{\texttt{optimization\_direction}}). Under \colorbox{Snow2}{\texttt{dsm\_settings}} the name of the direct search method (dsm) is given by \colorbox{Snow2}{\texttt{dsm\_algorithm\_name}} (e.g. ``NelderMead'') in the \colorbox{Snow2}{\texttt{general\_settings}} and stopping criteria are specified in \colorbox{Snow2}{\texttt{stopping\_criteria}} (see Sec.~\ref{sec:stopping_criteria}). Via \colorbox{Snow2}{\texttt{random\_number\_generator}} a seed can be defined to perform reproducible searches during testing, debugging, or determination of suitable hyperparameters. In this context, the hyperparameters are the settings in the configuration that have a direct influence on the optimization, such as the number of (super-) iterations, the specific basis used for pulse expansion, or the number of basis functions per search during a super-iteration. The stopping criteria and convergence tolerances can also be considered as hyperparameters because they determine the behavior of the optimization.

In experiments, specific issues may arise, such as the need to deal with measurement uncertainty and noise or drift of the detection signal. For simple drift compensation, a regular time interval on which to re-calibrate the current best set of pulses and parameters to the experiment can be specified in \colorbox{Snow2}{\texttt{compensate\_drift}} either as a time-periodic check or after each super-iteration.
Noisy measurements might require the re-evaluation of data points to ensure an accurate interpretation of the parameter landscape if the FoM evaluated for two data points are within the measurement error. The \colorbox{Snow2}{\texttt{re\_evaluation}} option uses a list of threshold probabilities and the standard deviation of the FoM as input. A potential improvement of the pulses is verified by repeating the evaluation to lower the measurement uncertainty according to the criteria specified by the threshold probabilities. If this option is activated, a standard deviation of the measured FoM, or at least a reasonable estimate, has to be provided in the \texttt{AbstractFoM} class.

Custom designs of search algorithms can be written according to the available templates or by adapting existing scripts coming with QuOCS. The Python files containing the algorithm class (see Sec.~\ref{sec:dsm}) can be placed in any folder of the user's file structure and linked via the keys \colorbox{Snow2}{\texttt{dsm\_algorithm\_module}} for the filename and \colorbox{Snow2}{\texttt{dsm\_algorithm\_class}} for the name of the class so that QuOCS can find and use them.

\subsubsection{Configuration of Pulses and Parameters}
\label{sec:pulses_and_params}
Time-dependent control functions can be defined under \colorbox{Snow2}{\texttt{pulses}}. \colorbox{Snow2}{\texttt{pulse\_name}} allows the user to provide a custom name, e.g. "amplitude" or "voltage", and the associated duration is connected by \colorbox{Snow2}{\texttt{time\_name}}. This feature enables the optimization of several pulses on different time scales and to specify the same duration for several pulses. Other parameters for pulses are the amplitude limits (\colorbox{Snow2}{\texttt{upper\_limit}} and \colorbox{Snow2}{\texttt{lower\_limit}}), the number of bins into which to discretize the pulse (typically the sampling rate of the electronics in the experiment or the model resolution) by \colorbox{Snow2}{\texttt{bins\_number}} and an initial amplitude variation scale via \colorbox{Snow2}{\texttt{amplitude\_variation}}. The amplitude variation is specified in the units of the amplitude limits and guess pulse (if provided) and describes the variation of the points in the start simplex in case of Nelder-Mead or the initial width of the Gaussian in case of CMA-ES. Initial guesses can be defined by providing a Python lambda function or a list of values under \colorbox{Snow2}{\texttt{initial\_guess}} that can, e.g., stem from previous optimization runs. This list can also be added programmatically to the settings dictionary before executing the optimization by reading in the results from a previous run or sampling from a function defined elsewhere. The same holds for a scaling function, which can be used to set the pulse amplitudes to zero at the beginning and the end or to include post-processing of the obtained pulse-shape. In the \colorbox{Snow2}{\texttt{basis}} section, the desired basis in which to expand a pulse can be set and further parameters that are specific to the chosen basis can be defined. These include the distribution (\colorbox{Snow2}{\texttt{distribution\_name}}) from which to select the random variables for the dCRAB algorithm (see \ref{app:dCRAB}), the number of basis vectors per super-iteration (\colorbox{Snow2}{\texttt{basis\_vector\_number}}), etc.. QuOCS comes equipped with a set of bases: Fourier, Chebyshev, piece-wise constant, Walsh, and Sigmoid.  However, the user can easily extend this list by writing their own basis according to the available template and link their file to QuOCS in the basis part of the pulse settings. As for the custom algorithm, the Python files containing the new basis class can be placed in any folder of the user's file structure and referenced via the keys \colorbox{Snow2}{\texttt{basis\_module}} for the filename and \colorbox{Snow2}{\texttt{basis\_class}} for the name of the class so that QuOCS can find and use them.

Under \colorbox{Snow2}{\texttt{times}} the settings so far contain a name to connect it to a pulse, an initial value for the duration and ranges in which the duration might be varied with a reasonable variation size to start with. Currently, the variation and optimization of pulse durations is not supported yet.

The same as for the \colorbox{Snow2}{\texttt{times}} holds for \colorbox{Snow2}{\texttt{parameters}}, except that they should be seen as constant but variable values that are optimized besides the pulse amplitudes and are not connected to the pulses (at least not inside QuOCS), e.g. the constant detuning of a control field or a bias voltage. Here, the variation is considered in the optimization and the final values are saved with the optimal pulses (see Sec.~\ref{sec:the_optimization}).

\subsubsection{Stopping Criteria}
\label{sec:stopping_criteria}
Numerous stopping criteria are available to either end the optimization upon convergence or continue with a new super-iteration if a local search is stuck. Globally, one can set the total number of function evaluations by \colorbox{Snow2}{\texttt{max\_eval\_total}}, the total time after which to stop by \colorbox{Snow2}{\texttt{total\_time\_lim}} given in minutes, and a goal FoM that is to be reached by \colorbox{Snow2}{\texttt{FoM\_goal}} given in the same units as the FoM provided by the \texttt{AbstractFoM} class. These options are defined directly in the algorithm settings.

For stopping criteria of individual super-iterations (SI) a \colorbox{Snow2}{\texttt{stopping\_criteria}} entry in the direct search method settings is used. Among them are convergence criteria for the search method defined analogous to numpy like \colorbox{Snow2}{\texttt{xatol}} and, \colorbox{Snow2}{\texttt{frtol}} which describe the simplex size and relative variation of the FoM over the simplex points, respectively, in the case if Nelder-Mead. Additionally, a time limit for the SI, a maximum number of evaluations, and a change-based stop can be specified. The change-based stop takes the slope of the FoM trend over a given number of evaluations, below which the optimization proceeds with the next SI. This is a very intuitive way to save time in an already sufficiently converged local search.

\subsection{The Optimizer Class}
\label{sec:the_optimization}

After the configuration of the optimization problem, the iterative optimization process can start.

\begin{figure}[!htb]
 \centering
 \includegraphics[scale=0.9]{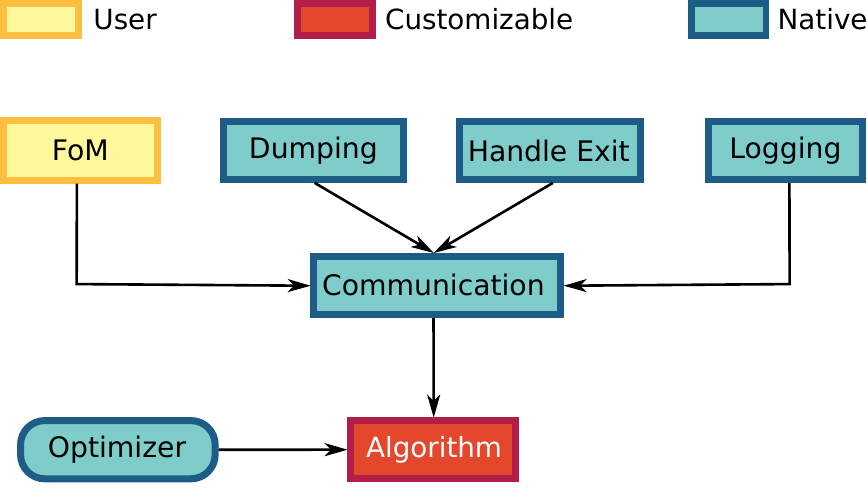}
 \caption{\textbf{General structure of QuOCS.} The \texttt{FoM} class, which is specified by the user, is fed into a communication object together with further utilities for data dumping, logging and exit handling. The optimizer then connects this object to the defined optimization algorithm and initiates the run.}
 \label{fig:quocs_general_structure}
\end{figure}

The \texttt{Optimizer} class is the main class of the optimizer. As shown in Fig.~\ref{fig:quocs_general_structure}, it contains all the general modules every optimization algorithm requires, such as an object for the communication, which takes care of the data management (dumping), logging, and the current state of the optimization. Exit handling is also performed in here. One important job of the communication is to take care of the connection to a graphical interface which is currently under development. Also, the visualization and initiation of a run in the lab-software Qudi is enabled by this class. 


Each execution of the optimizer will create a folder named ``QuOCS\_Results'' in the current directory of the Python shell with which the script is called. The results of an optimization run are put in a sub-folder therein named with a date- and time-stamp plus the custom optimization name defined by the user. In that folder, all the results are saved. This contains a log file that can be used for investigations of errors during an optimization and to gain more insight in the optimization history. The software generates a \lstinline{.npz} file containing the best control pulses and parameters, the best achieved FoM and more information like the SI and iteration number required to converge. Lastly, a copy of the settings dictionary in JSON format and a text file with the used QuOCS version number is saved. The files that are saved can be extended manually by extracting desired information from the optimizer object or by adding functions to the FoM class.\\





\subsection{Optimization Algorithms}
\label{sec:optimization_algorithm}

\begin{figure}[!htb]
 \centering
 \includegraphics[scale=0.9]{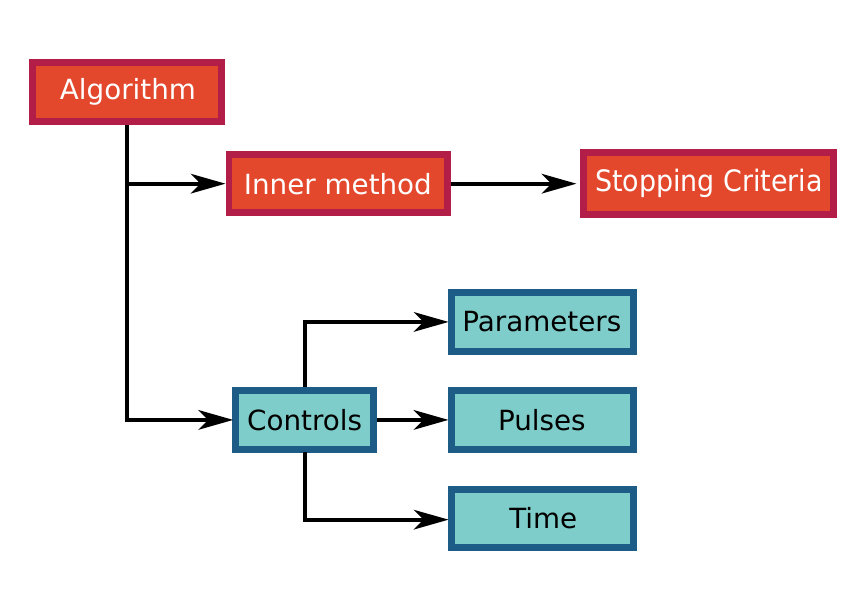}
 \caption{\textbf{The algorithm structure of QuOCS.} The inner method contains the main part of the algorithm and can be adapted by the user or replaced by a custom code. It is connected to a stopping criteria object to check for, e.g., convergence or timeout. The controls object handles the composition of pulses, constant but variable parameters and times and can be accessed by the optimization algorithm.}
 \label{fig:quocs_algorithm_structure}
\end{figure}

The optimization code is at the core of the Quantum Optimal Control Suite.
The chosen optimization algorithm itself is a child class of \texttt{OptimizationAlgorithm} and invokes all independent module classes which it needs to perform the optimization as well as information about the communication object. A visualization of the structureis shown in Fig.~\ref{fig:quocs_algorithm_structure}.

A routine defined in the \texttt{OptimizationAlgorithm} class is called every time a set of controls is evaluated. At every call of the routine, the global stopping criteria are checked and the communication with the linked FoM class is established. The child class for a specific algorithm has an inner routine call that is linked to its parent class and function. In the child, the specific stopping criteria are checked, and the re-evaluation steps are performed. 

The algorithm itself contains an object for the controls (pulses, parameters and times), so that the used basis and settings can be handled separately, and the optimization algorithms are kept general. Also, the stopping criteria are taken care of in their own class, so that for each search strategy unique convergence indicators can be considered. During the routine call, only a query if one of the criteria has been reached is done and acted upon accordingly.

\subsection{Direct search methods}
\label{sec:dsm}

During each dCRAB super-iteration, a direct search for the set of randomized controls is performed. The \texttt{DirectSearchMethod} class implements different direct search algorithms, such as Nelder-Mead and CMA-ES. Further strategies, that each have different advantages depending on the control problem, can easily be added personalized by the user. Direct search methods that are written or adapted by a user can be linked in the JSON configuration file, as already explained for the user \texttt{OptimizationAlgorithm} and \texttt{basis}. In this way, the new code can be integrated in the local file structure of the user, without the need to update the QuOCS installation (see Sec. \ref{sec:Configuration}).

\subsection{Bases and Pulses}
\label{sec:bases_and_pulses}

\begin{figure}[!htb]
 \centering
 \includegraphics[scale=0.9]{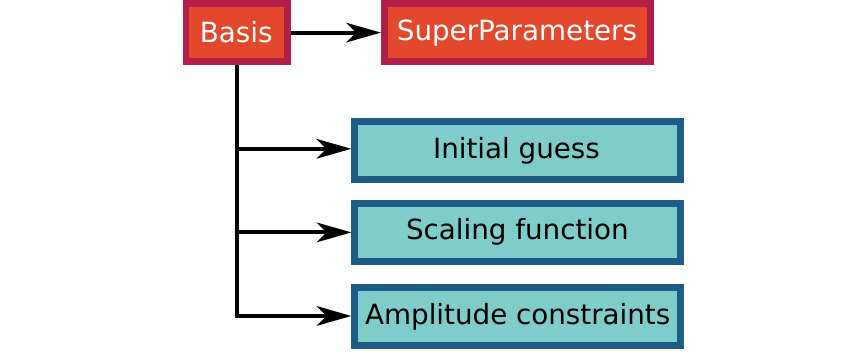}
 \caption{\textbf{Structure of pulses in QuOCS.} The basis class obtains the information about each pulse, such as the function space in which to expand it and the optimization-parameters used for optimization. An initial guess, a scaling function and possible amplitude constraints are also defined here.}
 \label{fig:quocs_pulses_structure}
\end{figure}

Since the time-dependent pulses are expanded in some basis, a separate class for the pulses has been implemented. Each basis type is a child of this class and combines the to be optimized optimization-parameters to describe a certain pulse shape depending on the specified basis and the initial guess provided by the user. Here, also the scaling function is used to post-shape the pulse if needed. Finally, amplitude constraints are enforced. The structure can be visualized in Fig.~\ref{fig:quocs_pulses_structure}. Several ways are possible here: simply cutting off the parts of a pulse that exceed the limits can introduce sharp edges, which can be useful to achieve shapes similar to bang-bang controls \cite{hoeb2017amplification}. If bandwidth restrictions are important for the setup or system at hand, the shrink option can be useful, where the pulse is squeezed until it obeys all limitations. Additionally, post-processing steps to enforce bandwidth-limitations by cutting off high or low frequencies after a Fourier transformation could be implemented.

\section{Example Applications}
\label{sec:example_applications}



In the following subsections, we are going to show three optimization examples to illustrate how QuOCS can be used to improve processes using models and experimental feedback. One of the model-based optimizations is demonstrated with a gradient-based and one with a gradient-free algorithm. Lastly, we show the results of a simple closed-loop optimization on an experiment in a laboratory.

\subsection{Model-based Optimization with GRAPE}
\label{sec:GRAPE_example}

We show the performance of GRAPE on an Ising problem with a Hamiltonian similar to the one used in Ref.~\cite{Felix_deep_learning_2022}. The optimization problem is given by
\begin{equation}
\label{eq:Hamil_ising_chain}
H(t) = -J \sum_j \sigma_j^z \sigma_{j+1}^z \; - g \sum_j \sigma_j^z \sigma_{j+2}^z \; + u(t) \sum_j \sigma_j^x
\end{equation}
where $J$ and $g$ describe the nearest and next-nearest neighbor interaction, respectively. A global control field in the x-direction $u(t)$ is used for the control of the system. We use a spin-chain of 5 qubits and take $J=1$ and $g=2$. A pulse of length 1 (a.u.) is discretized into 100 piece-wise constant elements. 
In QuOCS, we exploit the GRAPE algorithm to optimize the control field $u(t)$ from an initial guess of $u(t) = 0$ to find an operation that transfers all spins from the ground to the excited state.
The figure of merit (FoM) is defined as the fidelity of the overlap of the target state with the state after time evolution under the control pulse according to the Hilbert-Schmidt inner product $F = \text{tr}[\rho(T)^{\dagger} \rho_{\text{aim}}]$. The time propagation is performed by a piece-wise evolution method using the matrix-exponential provided by QuOCS. 


Initially, the modules needed for setting up the QOC problem are imported.

\begin{lstlisting}[language=Python]x
from quocslib.utils.inputoutput import readjson
from quocslib.utils.AbstractFoM import AbstractFoM
from quocslib.Optimizer import Optimizer
\end{lstlisting}

To define the control problem, the user provides a class which handles the dynamics of the system. At this point the drift- and control-Hamiltonian might be provided or the figure of merit (FoM) is handed back directly establishing the connection to the experiment or simulation.

\begin{lstlisting}[language=Python]
class OneQubit(AbstractFoM):

    def __init__(self, args_dict:dict = None):
        """ Initialize the dynamics variables"""
        if args_dict is None:
            args_dict = {}
        ...

    def get_FoM(self, pulses: list = [],
                parameters: list = [],
                timegrids: list = []
        ) -> dict:
        # Compute the dynamics and FoM
        ...

        return {"FoM": fidelity}

# Create Figure of Merit object
FoM_object = OneQubit()
\end{lstlisting}

The optimization algorithm is customized via a JSON file that contains information about the number of pulses, stopping criteria, etc..

\begin{lstlisting}[language=Python]
optimization_dictionary = readjson("opt_dictionary.json"))
\end{lstlisting}

The JSON dictionary used for the GRAPE optimization is as follows:

\begin{lstlisting}[language=Python]
{
    "algorithm_settings": {
        "algorithm_name": "GRAPE",
        "stopping_criteria": {
            "max_eval_total": 100,
            "ftol": 1e-6,
            "gtol": 1e-6
        }
    },
    "pulses": [{
        "pulse_name": "Pulse_1",
        "upper_limit": 100.0,
        "lower_limit": -100.0,
        "bins_number": 100,
        "amplitude_variation": 20.0,
        "time_name": "time_1",
        "basis": {
            "basis_name": "PiecewiseBasis",
            "bins_number": 100
        },
        "scaling_function": {
            "function_type": "lambda_function",
            "lambda_function": "lambda t: 1.0 + 0.0*t"
        },
        "initial_guess": {
            "function_type": "lambda_function",
            "lambda_function": "lambda t: 0.0 + 0.0*t"
        }
    }],
    "parameters": [],
    "times": [{
        "time_name": "time_1",
        "initial_value": 1.0
    }]
}
\end{lstlisting}

In the last step, the optimizer object is built from the provided class and dictionary. After that the QuOCS optimization can be started:

\begin{lstlisting}[language=Python]
# Define Optimizer
optimization_obj = Optimizer(optimization_dictionary, FoM_object)

# Execute the optimization
optimization_obj.execute()
\end{lstlisting}

\begin{figure}[!htb]
 \centering
 \includegraphics[width=0.45\textwidth]{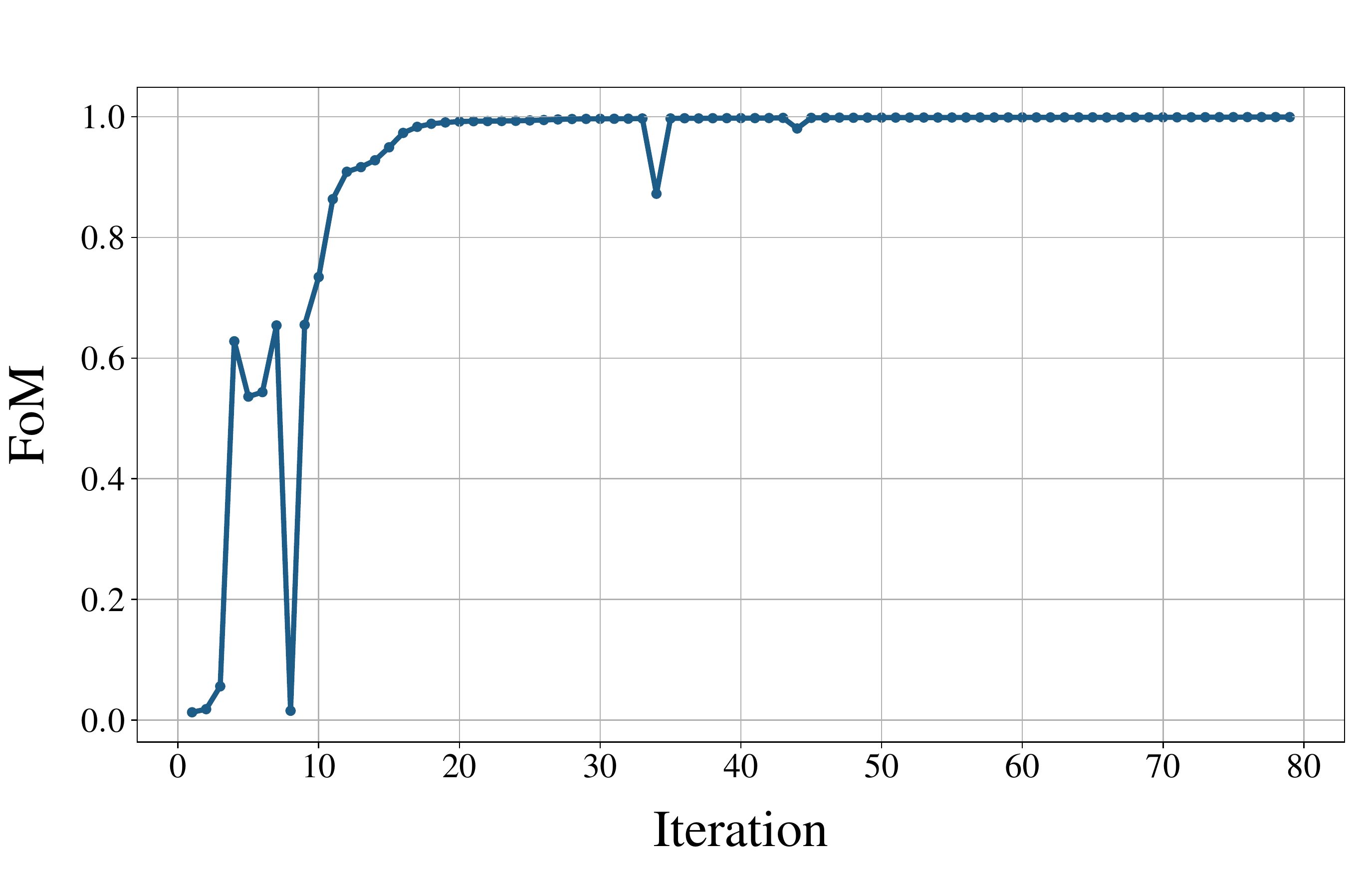}
 \includegraphics[width=0.45\textwidth]{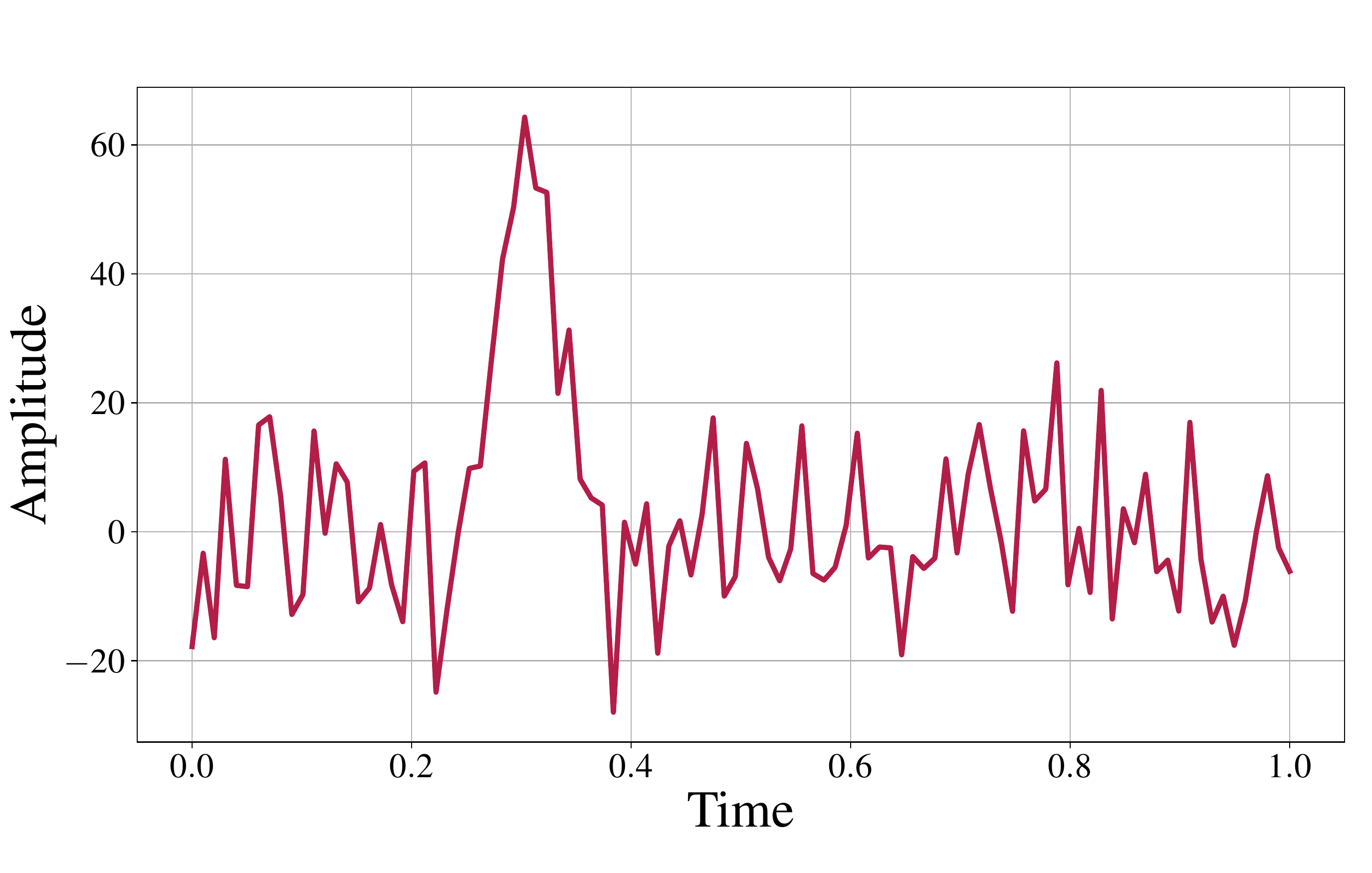}
 \caption{\textbf{Optimization results with GRAPE:} \textit{Top:} Evolution of the figure of merit during the iterations of the GRAPE algorithm for the system described by the Hamiltonian in eq. (\ref{eq:Hamil_ising_chain}). \textit{Bottom:} Final optimized amplitude $u(t)$ after the application of GRAPE on the problem.}
 \label{fig:GRAPE_opt}
\end{figure}


The final optimized amplitude and evolution of the FoM during the run is shown in Fig.~\ref{fig:GRAPE_opt} and the fidelity reached a value of 99.95\% in 79 steps of the GRAPE algorithm. This can be pushed further by decreasing the convergence tolerances (\colorbox{Snow2}{\texttt{ftol}} and \colorbox{Snow2}{\texttt{gtol}}) of the L-BFGS-B minimizer used during the optimization~\cite{SciPy}.\\

\subsection{Model-based Optimization with dCRAB}
\label{sec:dCRAB_example}

Now, we define the same model of the previous section but the optimization is performed via the dCRAB algorithm to give an example of a gradient-free optimization. 
However, to slightly modify the problem and to emulate fluctuations of the output of the system, a variation of the next-nearest neighbor coupling is added to the simulation as an example. 
%
%
For this purpose, we sample a random number from a Gaussian distribution around 0 with a variance of 0.1 and add it to the value of $g$ in each evaluation of the FoM. In such a "noisy" scenario the dCRAB algorithm, in combination with Nelder-Mead and a Fourier basis under usage of the re-evaluation steps option of QuOCS, is a good approach to solve the problem. One reason for this is the natural robustness of Nelder-Mead since its simplex size can be chosen larger than the variations in the parameter landscape. The Fourier basis can produce arbitrary pulse shapes limited by the bandwidth provided in the optimization settings and the re-evaluations option is designed to consider uncertainty in the evaluated FoM. The settings for the optimization were defined as follows:

\begin{lstlisting}[language=Python]
{
    "optimization_client_name": "Optimization_dCRAB_IsingModel",
    "algorithm_settings": {
        "algorithm_name": "dCRAB",
        "super_iteration_number": 3,
        "max_eval_total": 2000,
        "dsm_settings": {
            "general_settings": {
                "dsm_algorithm_name": "NelderMead",
                "is_adaptive": true
            },
            "stopping_criteria": {
                "xatol": 1e-14,
                "frtol": 1e-3,
                "change_based_stop": {
                    "cbs_funct_evals": 200,
                    "cbs_change": 0.01
                }
            }
        },
        "re_evaluation": {}
    },
    "pulses": [
        {
            "pulse_name": "Pulse1",
            "upper_limit": 1000.0,
            "lower_limit": -1000.0,
            "time_name": "time1",
            "amplitude_variation": 10.0,
            "basis": {
                "basis_name": "Fourier",
                "basis_vector_number": 5,
                "random_super_parameter_distribution": {
                    "distribution_name": "Uniform",
                    "lower_limit": 0.01,
                    "upper_limit": 5.0
                }
            },
            "scaling_function": {
                "function_type": "lambda_function",
                "lambda_function": "lambda t: 1.0 + 0.0*t"
            },
            "initial_guess": {
                "function_type": "lambda_function",
                "lambda_function": "lambda t: 0.0 + 0.0*t"
            }
        }
    ],
    "times": [
        {
            "time_name": "time1"
        }
    ],
    "parameters": [],
    "communication": {
        "communication_type": "AllInOneCommunication",
        "results_folder": "/home/thomas/sciebo/PhD/RedCRAB/QuOCS/QuOCS_Results"
    }
}
\end{lstlisting}

\begin{figure}[!htb]
 \centering
 \includegraphics[width=0.45\textwidth]{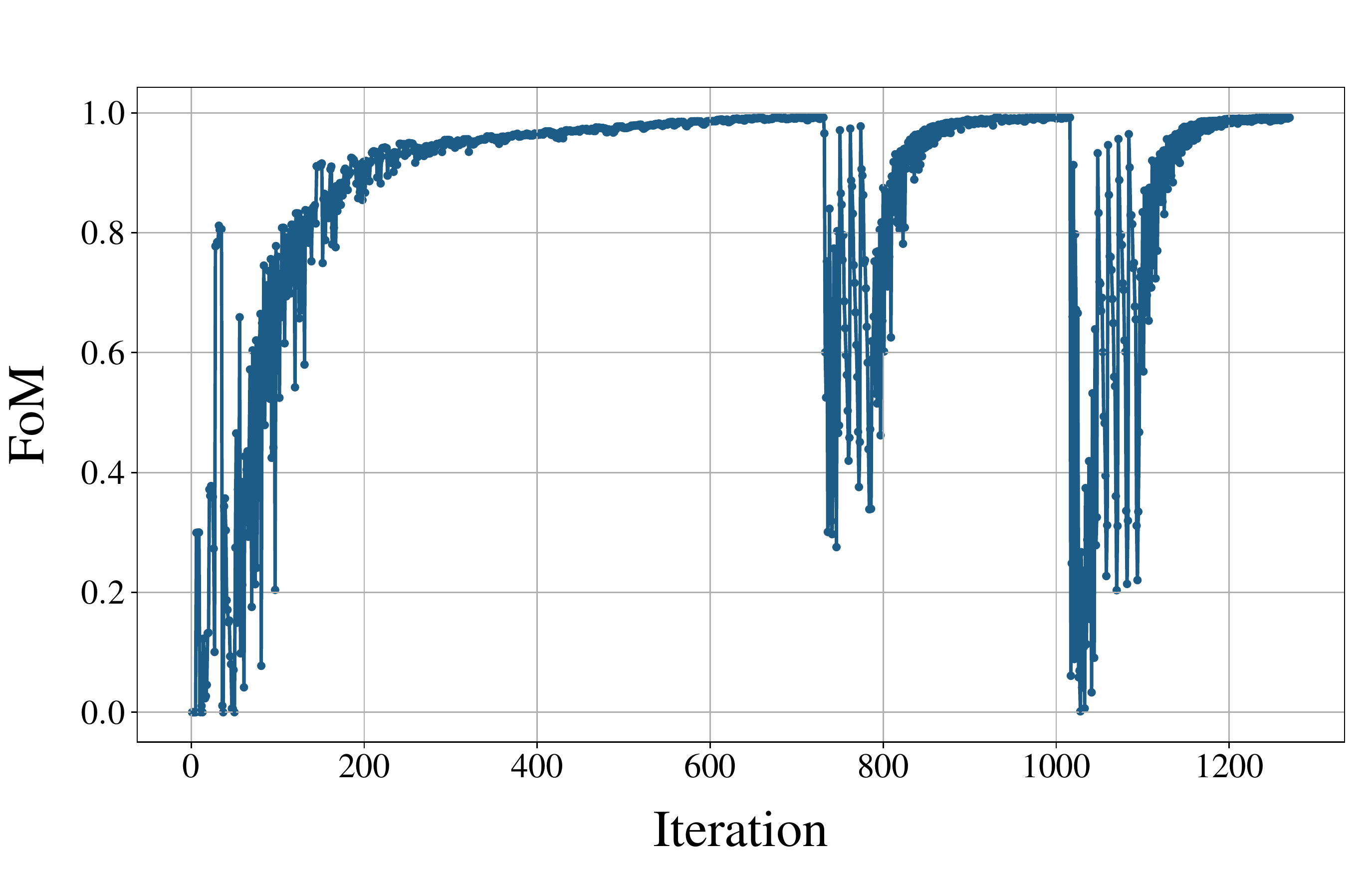}
 \includegraphics[width=0.45\textwidth]{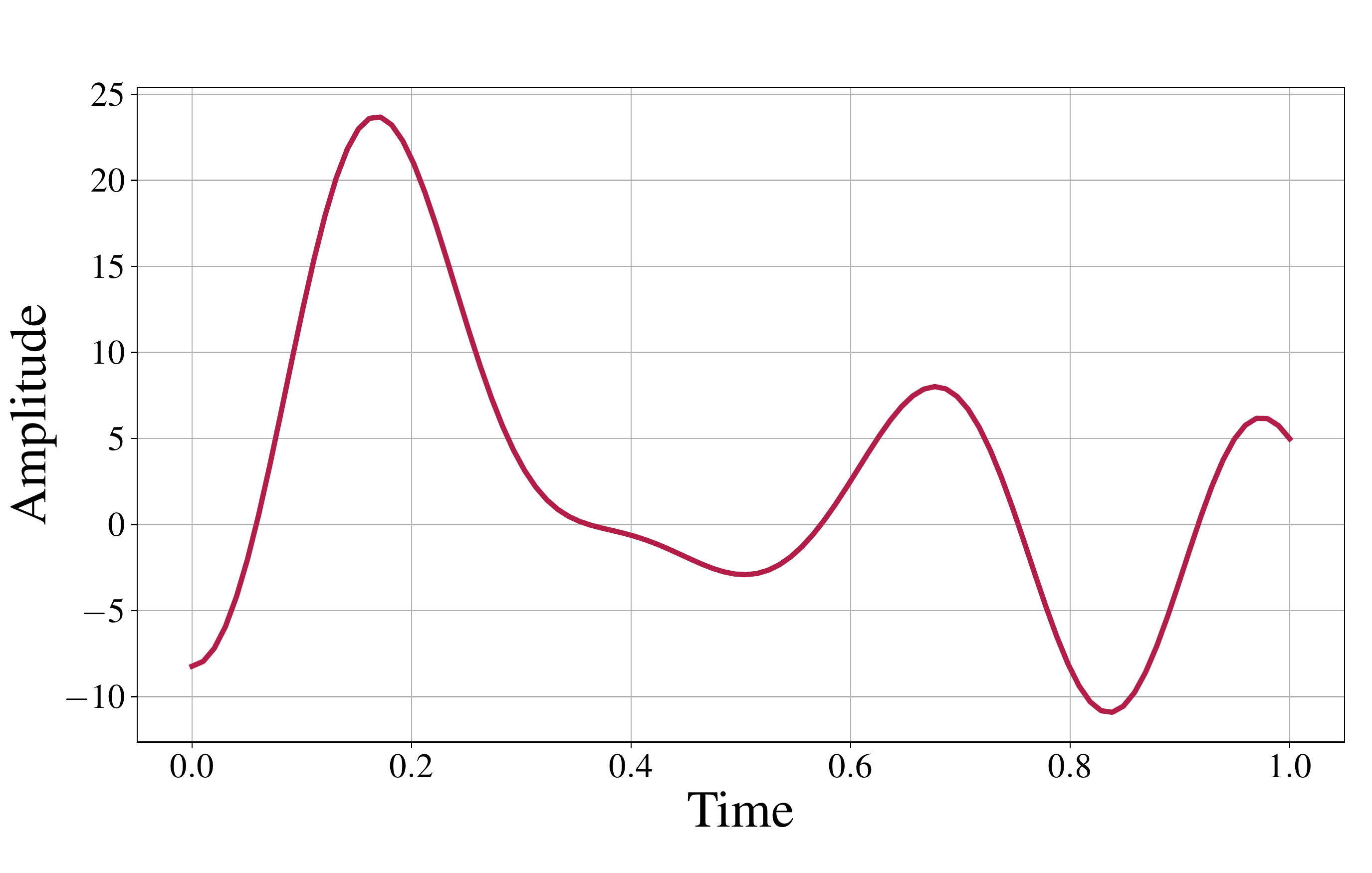}
 \caption{\textbf{Optimization results with dCRAB:} \textit{Top:} Evolution of the figure of merit during the iterations of the dCRAB algorithm for the system described by the Hamiltonian in eq. (\ref{eq:Hamil_ising_chain}) with additional noise on the $g$ parameter. \textit{Bottom:} Final optimized amplitude $u(t)$ after the application of dCRAB on the problem.}
 \label{fig:dCRAB_opt}
\end{figure}

In Fig.~\ref{fig:dCRAB_opt} the evolution of the FoM during an optimization with the dCRAB algorithm and the resulting control is shown. At step 730 and 1015 a new dCRAB super-iteration is started and the new parameter landscape is searched by the NM method with an (initially) larger simplex (see \cite{rachDressingChoppedrandombasisOptimization2015a, heck2018remote, Rembold2020}). The final FoM results in 99.229$\pm$0.008\% for an average of the FoM evaluated 50 times with the best pulse provided after the run.

The resulting pulse is much smoother than the GRAPE result. This feature is inherently provided by the dCRAB method and can be added to GRAPE as well by extending this basic implementation example. For instance, this can be done by parameterization of the pulse similar as in dCRAB and subsequent optimization of the coefficients by looking at their gradient as it is done with gradient optimization using parameterization (GROUP)~\cite{Soerensen_2018} or gradient optimization of analytic controls (GOAT)~\cite{Machnes_2018}. Another method is the usage of filter functions or interpolation~\cite{Motzoi_2011}. These features are candidates for potential future additions to the codebase of QuOCS.


\subsection{Closed-loop Implementation in an Experiment}
\label{sec:implementation_in_exp}

\begin{figure*}[!htb]
 \centering
\includegraphics[width=0.9\textwidth]{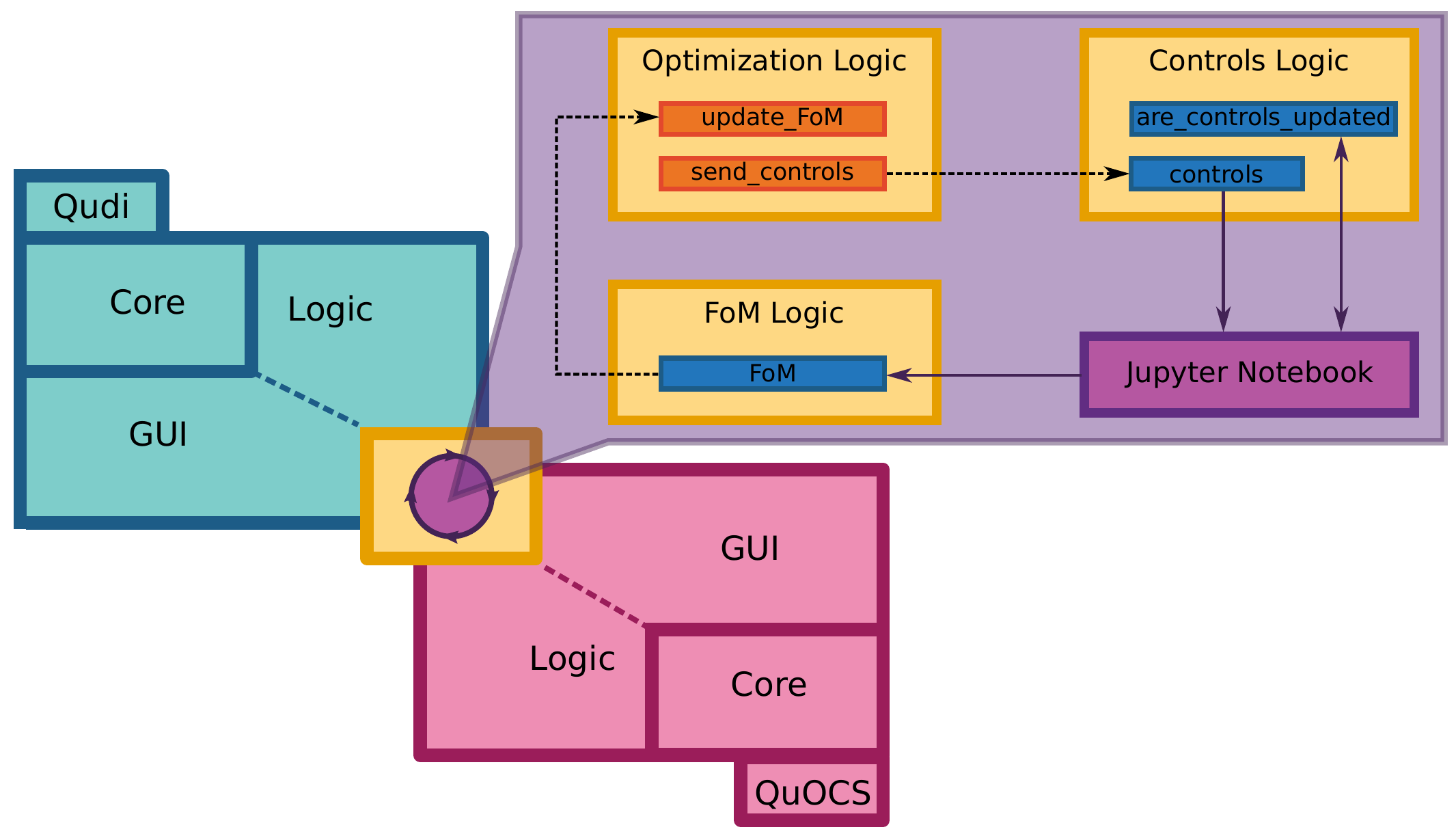}
 \caption{QuOCS implementation in Qudi.
 The \textit{optimization logic} class sends the controls to the Controls logic via a signal. The Jupyter notebook is then notified by the class variables \textit{are\_controls\_updated} when the controls are ready to be used in the measurement. Indeed the Qudi module inside the notebook performs the experiment and returns the feedback to the \textit{fom} variables in the \textit{fom\_logic} class. Finally, the feedback result is given back to the \textit{optimization\_logic} class via a signal and stored for the optimization algorithm. The red boxes delimit the class functions $($orange boxes$)$, and variables $($blue boxes$)$. The dashed arrows show the signals action among the logic classes, meanwhile the violet continuous lines presents the declaration and the return interactions between modules.}
 \label{fig:quocs_implementation_qudi_zoom}
\end{figure*}

To improve the usability of QuOCS in an experimental environment, the control suite contains an interface with Qudi~\cite{Binder2017}, a well-established multi-modular software to tune technologies often used for color-centers in diamond, as well as on other setups for experiments on quantum systems. The communication between the two platforms is established by using the signal slot feature of the Qt library and a Jupyter notebook kernel provided by Qudi. 
The optimization is started via a Jupyter notebook, controlling the status of QuOCS and Qudi at each iteration. A schematic of the process is provided in Fig.~\ref{fig:quocs_implementation_qudi_zoom}.\\


To test and verify the applicability of QuOCS in a real experiment, we perform a state-to-state transfer for an ensemble of nitrogen vacancy (NV) centers in diamond.
The NV center is a defect in the diamond lattice consisting of a vacancy and a nitrogen atom on the neighboring lattice site \cite{DOHERTY20131,PhysRevLett.102.057403, PhysRevLett.121.060401}. 
Because of its large zero-field splitting of $D\approx2.87~\text{GHz}$, its $m_s=-1$ and $m_s=0$ ground states can be treated as an effective two-level system when a magnetic magnetic field is applied (here $B=458~\text{G}$). The applied magnetic field also leads to polarization of the nuclear spin associated with the nitrogen of the NV center~\cite{PhysRevLett.102.057403}.
The diamond sample is grown by chemical vapor deposition and has a natural abundance of ${}^{13}\text{C}$ and an NV layer thickness of $\approx 10~\mu\text{m}$ ([N] = 14 ppm).
Due to the large NV layer thickness and readout area of $\diameter \approx 14.5~\mu\text{m}$, the NV centers suffer from strong amplitude variation across the sample during the application of microwave pulses for spin state manipulation.
To overcome this problem, we task QuOCS to optimize the amplitude and phase of a 200 ns long pulse to perform a $\pi$ rotation of the NV spins from $m_s=0$ to $m_s=-1$.
As the $m_s=-1$ state shows a lower fluorescence signal than the $m_s=0$ state \cite{PhysRevB.81.035205}, the FoM is given by the fluorescence contrast between those two states.
We read out the NV centers spin state by applying a 532\,nm laser pulse, which can also be used to re-initialize it back to the $m_s=0$ state \cite{PhysRevB.74.104303}.
Generated pulses are sent in form of a microwave signal through a straight microwave antenna on top of the diamond. Further details of the setup can be found in~\ref{app:exp_setup}.
%
%

\begin{figure}[!htb]
 \centering
 \includegraphics[width=0.45\textwidth]{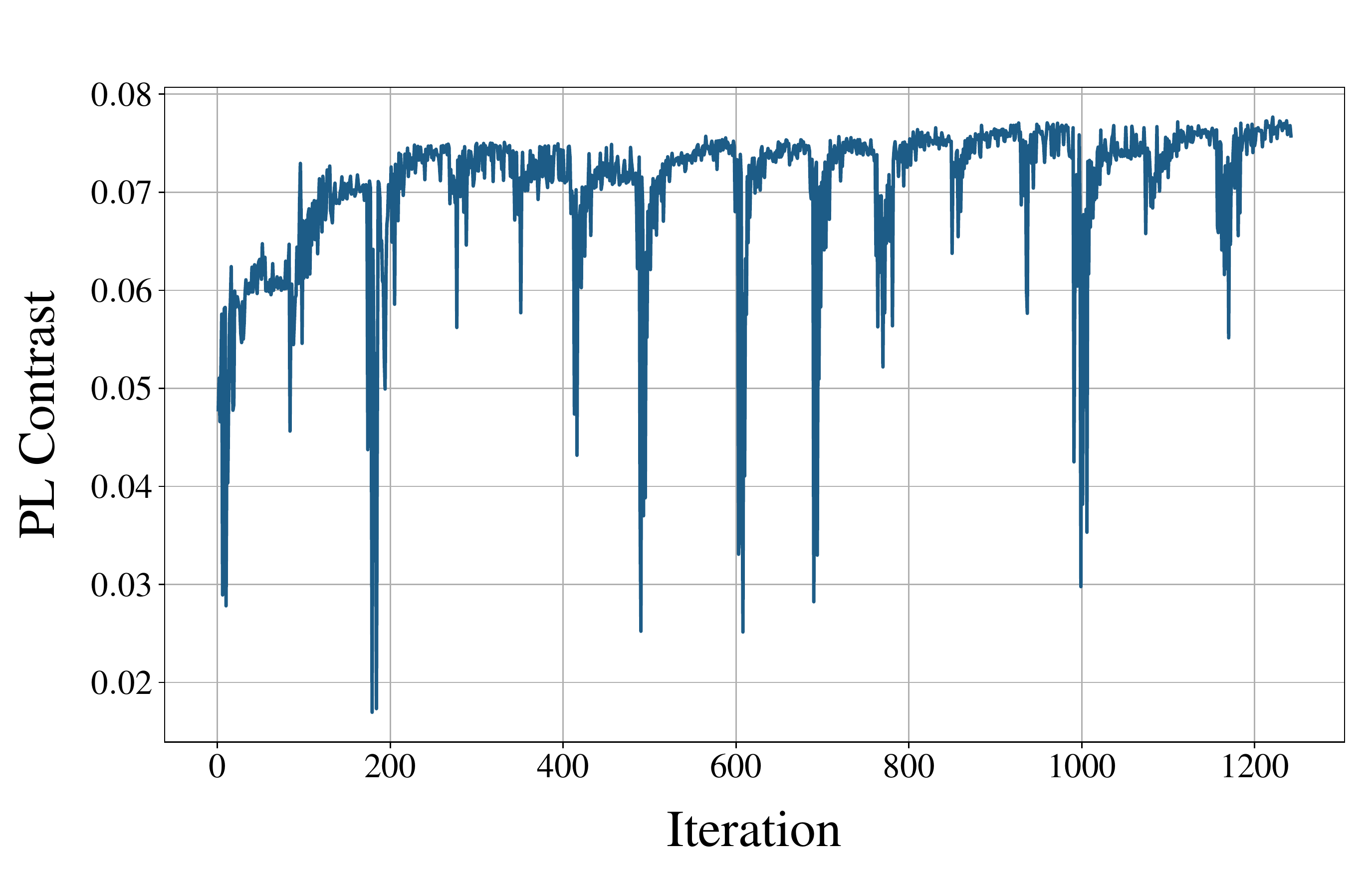}
 \includegraphics[width=0.45\textwidth]{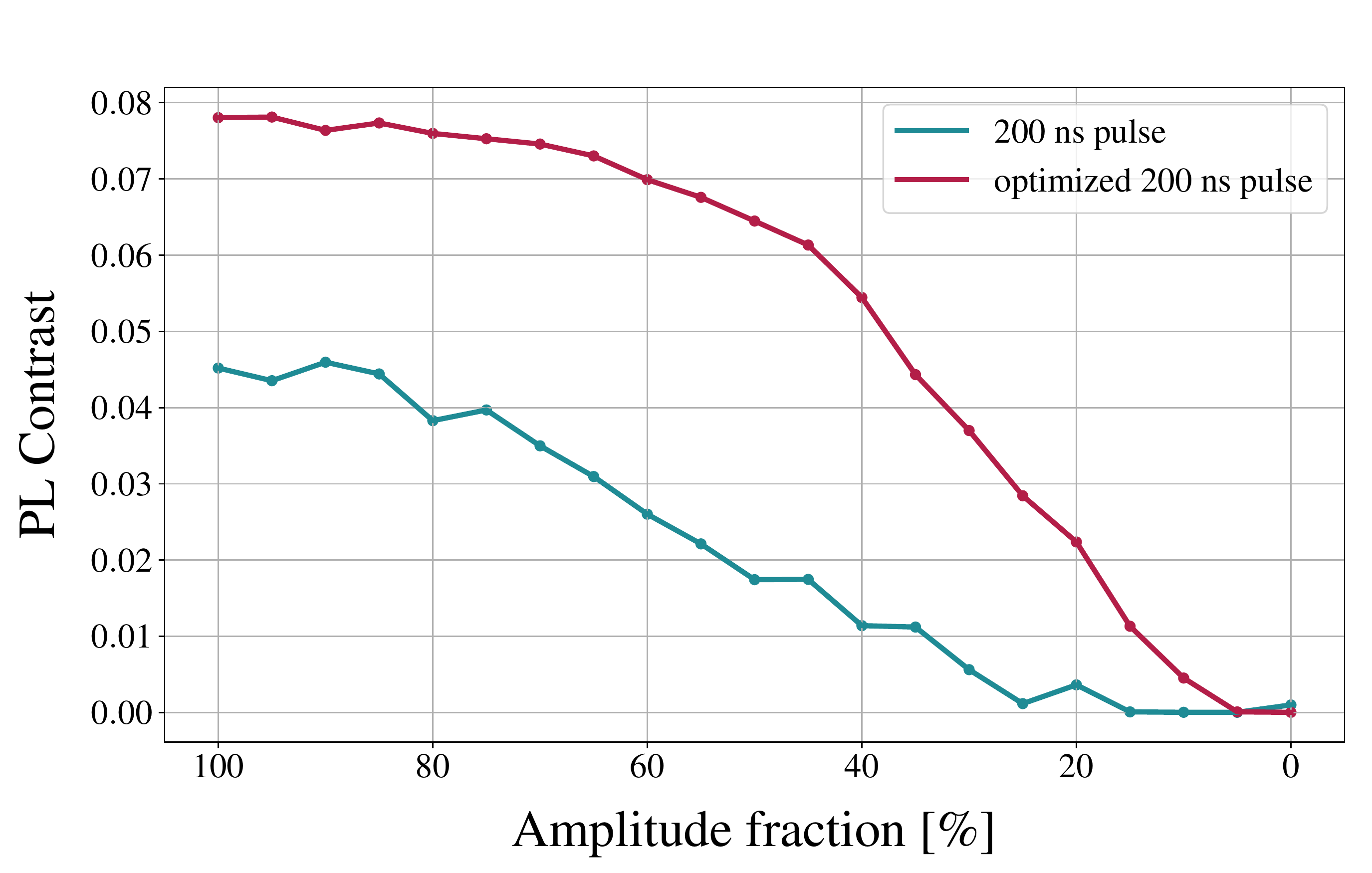}
 \caption{\textbf{Closed-loop optimization of a $\pi$-pulse on an NV center:} \textit{Top:} Evolution of the figure of merit during the iterations of the dCRAB algorithm for the maximization of the photo-luminescence (PL) contrast between the initial $m_s=0$ and excited $m_s=-1$ state. The PL contrast corresponds to the amount of population flipped from ground to excited state. \textit{Bottom:} Comparison of the achieved PL contrast between the optimized pulse and a simple rectangular pulse (constant amplitude and phase) under reduction of the amplitude from 100\% to 0.}
 \label{fig:experiment_opt}
\end{figure}

The optimization is performed for 3 hours where a new SI is started if the FoM does not change more than 0.001 over 50 evaluations.
One evaluation step takes $\sim$8.5\,s and is mainly limited by the time needed to generate the pulse and process the obtained fluorescence signal. This choice is mainly made to illustrate the behavior of the optimization over a longer timescale and several super-iterations. The improvement of the photoluminescence contrast is only minor after around 250 iterations and the optimization could already have been stopped there for a faster runtime of about 35 minutes.
As can be seen in the upper part of Fig.~\ref{fig:experiment_opt}, the FoM signal is subject to some drift over time, i.e. iterations. This drift is well mitigated by the compensation methods via a re-calibration of the current best pulse after every 15 minutes that is provided in QuOCS and taken into account for the final optimal pulse components. 
To investigate the robustness against amplitude errors, we compare the final optimized pulse to a rectangular one with equal length.
As shown in the lower part of Fig.~\ref{fig:experiment_opt}, the optimized pulse shows a much higher contrast in any case of artificially lowering the applied amplitude as compared to its rectangular counterpart and is therefore able to transfer more population from $m_s=0$ to $m_s=-1$.
%
%
Such optimized pulses can enhance the performance of experiments substantially \cite{vetter2021zero, frank2017autonomous, haberle2013high, dolde2014high, poulsen2022optimal} and due to its simple experimental implementation and fast optimization times QuOCS provides the perfect framework for these kinds of experiment.





\section{Conclusions and Outlook}
\label{sec:conclusions_and_outlook}

The family of QOC methods contains a variety of powerful algorithms to boost quantum technologies. To date, these methods have been implemented in several libraries, but a unified framework is missing. Finding the algorithm that is most suited to solve a problem, or having an heterogeneous approach, could be time demanding for a user in terms of the large amount of interfaces to develop for connecting the optimization algorithms to a given problem. Thus, to exploit the full potential of QOC and make these methods more accessible, even for researchers not in the QOC field, we provide QuOCS: a control suite that allows a problem to be solved by different algorithms in the same framework. Moreover, the modular structure of QuOCS allows an easier customization of the features to satisfy even specific needs of the user. For instance, new algorithms can be intuitively integrated with the many features already available and optimizer-related classes simplify the definition and implementation of new constraints, stopping criteria, etc.\\
We have shown with some examples that QuOCS simplifies also the process of transitioning from open-loop to closed-loop optimization. The open-loop optimization examples show the strength and weaknesses of two of the implemented algorithms, GRAPE and dCRAB. Instead, the closed-loop optimization is connected to the experiment via Qudi~\cite{Binder2017}, a modular experiment control software often used for experiments with color centers in diamond. While the experiment could also be connected using, e.g., a simple bash script we provide a module that allows the integration of QuOCS with the Qudi graphical user interface.\\

The field of quantum technology is in the transition phase between publicly funded research efforts and commercial exploitation, and QOC is starting to play a major role in this transition. Color centers in diamond, superconducting qubits, Rydberg atoms and ions in traps are vivid branches of quantum technology. Breakthroughs in these areas increasingly involve the need of tailored control over quantum devices and feedback loops between the device and the control algorithm. By providing and integration of QOC into the lab software Qudi, QuOCS will be a major step towards quantum technology based on color centers in diamond, especially with regard to automatic control of quantum computers and quantum sensors, and the modular structure allows for straightforward extension to many other platforms. However, any kind of setup being able to run in a closed-loop mode can take advantage by QuOCS.
Additionally - as demonstrated by similar open access control suites like QUTIP - QuOCS has the potential of widespread use across different quantum technology platforms. In particular, the underlying control algorithm dCRAB was used successfully on many different systems such as trapped ions, cold atoms, superconducting qubits or Bose Einstein Condensates~\cite{Mueller2022}. But also other fields of physics can profit from the application of QuOCS and OC in general~\cite{Angaroni_2019}.

Generally, QuOCS is available directly from a \href{https://github.com/Quantum-OCS/QuOCS}{GitHub} or PyPI repository. Installation and updating is hence straight-forward. Tutorials are available, as well as a list of QuOCS' features. Currently, we are working on the implementation of more updating algorithms, stopping criteria, and tutorials. To simplify the configuration of the optimization, which is submitted in the form of a JSON file, we are working on an adaptable GUI. Future updates might also include settings in the JSON file to automate often-used scenarios, such as automated visualization of pulses and FoM evolution.\\
As quantum technology is on the rise, so is the demand for QOC, applied to fully exploit the hardware's potential. QuOCS is designed to unify state-of-the-art QOC methods and provide a user-friendly open-source framework for quantum scientists and engineers.

\section{Conflict of Interest}

S.M. and T.C. are co-founders of Qruise GmbH. M.R. and A.M are employees of Qruise GmbH. All other authors declare no competing interests.




\section*{Acknowledgments}
\label{sec:acknowledgments}

This work was supported by the Helmholtz Validation Fund project ``Qruise” (HVF-00096), the European Union’s Horizon 2020 research and innovation program under the Marie Sk\l{}odowska-Curie \mbox{QuSCo} (Grant No. 765267) and the EU Quantum Flagship project \mbox{ASTERIQS} (Grant No. 820394). We acknowledge support from the Italian PRIN 2017, the EU project EURyQA, the EU-QuantERA projects QuantHEP and T-NISQ, and the WCRI-Quantum Computing and Simulation Center of Padova University. Furthermore, this project has received funding from the German Federal Ministry of Education and Research (BMBF) under the funding program quantum technologies $-$ from basic research to market $-$ with the QRydDemo grant. We also acknowledge funding by the BMBF via the projects SPINNING (No. 13N16210), VERTICONS (No. 13N14872) and QSolid. RSS and FJ acknowledge support from the DFG (CRC1279, EXC 2154 POLiS - Post Lithium Storage Cluster of Excellence, 445243414, 499424854), BMBF, BW Stiftung, Center for Integrated Quantum Science and Technology (IQst) and the European Research Council Synergy Grant HyperQ (Grant No. 319130). Finally, the research was supported by IQST and QuCoLiMa centers.

\appendix

\section{Optimal Control Algorithms}
\label{app:qoc_algorithms}

\subsection{Gradient-free algorithms}
\label{app:grad_free_algs}

QuOCS includes a set of direct search algorithms serving as a base for the dCRAB algorithm. 
The user has complete control over which algorithm to use and can tweak the hyper-parameters of the algorithm to enhance convergence.
Gradient-free optimization algorithms provide a recipe to find local minima in the absence of gradients. Gradients might not be available when the optimal control problem is especially complex or the access to data is limited. The latter is especially true for closed-loop optimizations where each FoM is obtained directly from the experiment. While finite-difference methods aim to approximate the gradient nevertheless, gradient-free optimizers are truly independent from them and commonly more efficient in the lower-dimensional searches we consider.
The software comes with a simplex-based algorithm (adaptive Nelder-Mead) as well as the Covariance Matrix Adaption Evolution Strategy (CMA-ES). However, the user can add their own gradient-free algorithms via the \texttt{DirectSearchMethod} class. An example implementation can be found under \texttt{GradientFreeTemplate.py}.
%
%
To reduce the number of parameters per optimization, gradient-free algorithms are commonly applied in combination with dCRAB.

\subsection{dCRAB}
\label{app:dCRAB}

The dressed Chopped RAndom Basis algorithm \cite{Doria_2011, canevaChoppedRandombasisQuantum2011a, rachDressingChoppedrandombasisOptimization2015a, Rembold2020} is designed to optimize time-dependent pulses by expanding the control field in a truncated basis of functions. Typical used bases are elements of the Fourier series, Chebyshev polynomials, Gaussian functions, or sigmoid functions with certain randomized parameters. The advantage of doing this is two-fold: experimental constraints such as, e.g., bandwidth limitations can be considered directly by limiting the parameters during basis vector selection. Secondly, the search space to find optimal update pulses is drastically lowered in comparison to optimizing the amplitudes of a piece-wise constant pulse because only the so-called optimization-parameters are varied and optimized. These optimization-parameters are the coefficients of the basis elements such as the amplitude and phase of a sine wave or the width of a Gaussian plus any constant but variable coefficients that need to be optimized as well. The basis vectors are randomized by selecting e.g. the frequency of the sine wave randomly from within a certain range. After finding the best-performing optimization-parameters for a set of basis functions, a new set of basis vectors is selected and the resulting pulse variations are added on top of the previous best solution. This scheme, split into several super-iterations, ensures that the improvement does not get stuck in a local optimum but can converge to the optimal solution given by the superordinate constraints (e.g. maximum amplitude, bandwidth, boundary conditions)~\cite{Li_Ruths_2009, Ruths_2011}.

\subsection{Gradient-based algorithms}
\label{app:grad_based_algs}

Gradient-based optimizations are implemented via GRAPE or automatic differentiation. As a basis QuOCS provides the gradient-based updating algorithms BFGS~\cite{nocedalNumericalOptimization2006} and LFBGS-B~\cite{byrdLIMITEDMEMORYALGORITHM, moralesRemarkAlgorithm7782011, zhuAlgorithm778LBFGSB1997} from Scipy~\cite{virtanenSciPyFundamentalAlgorithms2020} as well as a connection to the automatic differentiation library JAX~\cite{jax2018github}. 
Gradient-based optimization offers an efficient method of exploring high dimensional parameter spaces to find local optima. Gradient-based methods typically require prior knowledge of the task at hand so that an analytical gradient can be derived for use in the optimizer. If an expression cannot be found, then a finite difference approximation can be used. This is a numerical approach that approximates the gradient by evaluating it at $x$ and a small perturbation $x+\delta x$. Finite difference-based gradients are not included in this toolbox as their accuracy suffers from round-off error, and they are typically slow. Instead, for the supported problems, exact gradient methods are implemented. 
For gradient-based optimal control, we implement the Gradient Ascent Pulse Engineering algorithm (GRAPE) \cite{khaneja2005optimal,Machnes2011}. Time is discretized it into $N$ piecewise constant slices. During each slice, the system Hamiltonian, governing the dynamics, is assumed to be constant, which is generally a good approximation. The GRAPE algorithm included in this toolbox uses the ``auxiliary matrix method'' \cite{defouquieresSecondOrderGradient2011a, goodwinAdvancedOptimalControl2017, hogbenSpinachSoftwareLibrary2011, floetherRobustQuantumGates2012a} to compute the exact gradients of the system. State optimization and unitary gate synthesis are implemented by vectorizing the density matrix and performing the evolution in the Liouvillian space.
GRAPE performs well in high dimensional parameter spaces because it can exploit the gradient and make rapid progress. However, it offers no guards against becoming stuck in a local optimum within the parameter space. 
Also, included is our implementation of “AD-GRAPE'' a reverse-mode automatic differentiation-based version of GRAPE similar to the version implemented in \cite{PhysRevA.95.042318}. By making use of the JAX~\cite{jax2018github} framework, we are able to compile both the fidelity and gradient calculations functions and execute them rapidly, dramatically reducing the time it takes to optimize. We are also able to implement novel figures of merit where the analytical gradient is difficult or impossible to derive.

\section{Experimental Setup}
\label{app:exp_setup}

Once QuOCS forwards the optimized pulse shapes to Qudi, Qudi constructs the full measurement sequence. The measurement sequence is then uploaded to a Keysight M8195A arbitrary waveform generator (AWG) which offers a sample rate up to 65 GSa/s and generates the microwave signals. To decrease the upload speed and the overall optimization time, we lower the sample rate to 16.25 GSa/s. The microwave pulses are then send to an AR 30S1G6 amplifier (0.7-6 GHz, 30 W) and afterwards through the microwave antenna on the diamond surface (straight gold line on glass, produced by lithography).
The fluorescence counts are collected via a SiPM (Ketek PE3315-WB-TIA-SP) and the obtained voltage signal is recorded with a PCI Express Digitizer (Spectrum Instrumentation, M4i.4420-x8) and analyzed by Qudi.

\bibliographystyle{elsarticle-num}
\bibliography{refs}

\begin{thebibliography}{100}
\expandafter\ifx\csname url\endcsname\relax
  \def\url#1{\texttt{#1}}\fi
\expandafter\ifx\csname urlprefix\endcsname\relax\def\urlprefix{URL }\fi
\expandafter\ifx\csname href\endcsname\relax
  \def\href#1#2{#2} \def\path#1{#1}\fi

\bibitem{Brif_2010}
C.~Brif, R.~Chakrabarti, H.~Rabitz,
  \href{https://doi.org/10.1088/1367-2630/12/7/075008}{Control of quantum
  phenomena: past, present and future}, New Journal of Physics 12~(7) (2010)
  075008.
\newblock \href {https://doi.org/10.1088/1367-2630/12/7/075008}
  {\path{doi:10.1088/1367-2630/12/7/075008}}.
\newline\urlprefix\url{https://doi.org/10.1088/1367-2630/12/7/075008}

\bibitem{Glaser2015}
S.~J. Glaser, U.~Boscain, T.~Calarco, C.~P. Koch, W.~K{\"o}ckenberger,
  R.~Kosloff, I.~Kuprov, B.~Luy, S.~Schirmer, T.~Schulte-Herbr{\"u}ggen,
  D.~Sugny, F.~K. Wilhelm, Training schr{\"o}dinger's cat: quantum optimal
  control, The European Physical Journal D 69~(12) (2015) 279.
\newblock \href {https://doi.org/10.1140/epjd/e2015-60464-1}
  {\path{doi:10.1140/epjd/e2015-60464-1}}.

\bibitem{Koch_2016}
C.~P. Koch, \href{https://doi.org/10.1088/0953-8984/28/21/213001}{Controlling
  open quantum systems: tools, achievements, and limitations}, Journal of
  Physics: Condensed Matter 28~(21) (2016) 213001.
\newblock \href {https://doi.org/10.1088/0953-8984/28/21/213001}
  {\path{doi:10.1088/0953-8984/28/21/213001}}.
\newline\urlprefix\url{https://doi.org/10.1088/0953-8984/28/21/213001}

\bibitem{Koch_2022}
{Koch, Christiane P.}, {Boscain, Ugo}, {Calarco, Tommaso}, {Dirr, Gunther},
  {Filipp, Stefan}, {Glaser, Steffen J.}, {Kosloff, Ronnie}, {Montangero,
  Simone}, {Schulte-Herbr\"uggen, Thomas}, {Sugny, Dominique}, {Wilhelm, Frank
  K.}, \href{https://doi.org/10.1140/epjqt/s40507-022-00138-x}{Quantum optimal
  control in quantum technologies. strategic report on current status, visions
  and goals for research in europe}, EPJ Quantum Technol. 9~(1) (2022) 19.
\newblock \href {https://doi.org/10.1140/epjqt/s40507-022-00138-x}
  {\path{doi:10.1140/epjqt/s40507-022-00138-x}}.
\newline\urlprefix\url{https://doi.org/10.1140/epjqt/s40507-022-00138-x}

\bibitem{Mueller2022}
M.~M. Müller, R.~S. Said, F.~Jelezko, T.~Calarco, S.~Montangero,
  \href{https://doi.org/10.1088/1361-6633/ac723c}{One decade of quantum optimal
  control in the chopped random basis}, Reports on Progress in Physics 85~(7)
  (2022) 076001.
\newblock \href {https://doi.org/10.1088/1361-6633/ac723c}
  {\path{doi:10.1088/1361-6633/ac723c}}.
\newline\urlprefix\url{https://doi.org/10.1088/1361-6633/ac723c}

\bibitem{Rembold2020}
P.~Rembold, N.~Oshnik, M.~M. Müller, S.~Montangero, T.~Calarco, E.~Neu,
  Introduction to quantum optimal control for quantum sensing with
  nitrogen-vacancy centers in diamond, AVS Quantum Science 2~(2) (2020) 024701.
\newblock \href {http://arxiv.org/abs/https://doi.org/10.1116/5.0006785}
  {\path{arXiv:https://doi.org/10.1116/5.0006785}}, \href
  {https://doi.org/10.1116/5.0006785} {\path{doi:10.1116/5.0006785}}.

\bibitem{Gamkrelidze_1999}
R.~V. Gamkrelidze, \href{https://doi.org/10.1023/A:1021783020548}{Discovery of
  the maximum principle}, Journal of Dynamical and Control Systems 5~(4) (1999)
  437--451.
\newblock \href {https://doi.org/10.1023/A:1021783020548}
  {\path{doi:10.1023/A:1021783020548}}.
\newline\urlprefix\url{https://doi.org/10.1023/A:1021783020548}

\bibitem{Pesch_2012}
H.~J. Pesch, M.~Plail, The cold war and the maximum principle of optimal
  control, Documenta Mathematica (01 2012).

\bibitem{Geometric_Control_2002}
A.~Agrachev, Y.~Sachkov, Control Theory from the Geometric Viewpoint, 2002.
\newblock \href {https://doi.org/10.1007/978-3-662-06404-7}
  {\path{doi:10.1007/978-3-662-06404-7}}.

\bibitem{Clark_2021}
W.~Clark, M.~Oprea, A.~J. Graven, \href{https://arxiv.org/abs/2111.11645}{A
  geometric approach to optimal control of hybrid and impulsive systems}
  (2021).
\newblock \href {https://doi.org/10.48550/ARXIV.2111.11645}
  {\path{doi:10.48550/ARXIV.2111.11645}}.
\newline\urlprefix\url{https://arxiv.org/abs/2111.11645}

\bibitem{STA_1997}
R.~Unanyan, L.~Yatsenko, K.~Bergmann, B.~Shore,
  \href{https://www.sciencedirect.com/science/article/pii/S0030401897000990}{Laser-induced
  adiabatic atomic reorientation with control of diabatic losses}, Optics
  Communications 139~(1) (1997) 48--54.
\newblock \href {https://doi.org/https://doi.org/10.1016/S0030-4018(97)00099-0}
  {\path{doi:https://doi.org/10.1016/S0030-4018(97)00099-0}}.
\newline\urlprefix\url{https://www.sciencedirect.com/science/article/pii/S0030401897000990}

\bibitem{STA_2003}
M.~Demirplak, S.~A. Rice, \href{https://doi.org/10.1021/jp030708a}{Adiabatic
  population transfer with control fields}, The Journal of Physical Chemistry A
  107~(46) (2003) 9937--9945.
\newblock \href {http://arxiv.org/abs/https://doi.org/10.1021/jp030708a}
  {\path{arXiv:https://doi.org/10.1021/jp030708a}}, \href
  {https://doi.org/10.1021/jp030708a} {\path{doi:10.1021/jp030708a}}.
\newline\urlprefix\url{https://doi.org/10.1021/jp030708a}

\bibitem{STA_2010}
X.~Chen, A.~Ruschhaupt, S.~Schmidt, A.~del Campo, D.~Gu\'ery-Odelin, J.~G.
  Muga, \href{https://link.aps.org/doi/10.1103/PhysRevLett.104.063002}{Fast
  optimal frictionless atom cooling in harmonic traps: Shortcut to
  adiabaticity}, Phys. Rev. Lett. 104 (2010) 063002.
\newblock \href {https://doi.org/10.1103/PhysRevLett.104.063002}
  {\path{doi:10.1103/PhysRevLett.104.063002}}.
\newline\urlprefix\url{https://link.aps.org/doi/10.1103/PhysRevLett.104.063002}

\bibitem{STA_2013}
E.~Torrontegui, S.~Ibáñez, S.~Martínez-Garaot, M.~Modugno, A.~{del Campo},
  D.~Guéry-Odelin, A.~Ruschhaupt, X.~Chen, J.~G. Muga,
  \href{https://www.sciencedirect.com/science/article/pii/B9780124080904000025}{Chapter
  2 - shortcuts to adiabaticity}, in: E.~Arimondo, P.~R. Berman, C.~C. Lin
  (Eds.), Advances in Atomic, Molecular, and Optical Physics, Vol.~62 of
  Advances In Atomic, Molecular, and Optical Physics, Academic Press, 2013, pp.
  117--169.
\newblock \href
  {https://doi.org/https://doi.org/10.1016/B978-0-12-408090-4.00002-5}
  {\path{doi:https://doi.org/10.1016/B978-0-12-408090-4.00002-5}}.
\newline\urlprefix\url{https://www.sciencedirect.com/science/article/pii/B9780124080904000025}

\bibitem{STA_2019}
D.~Gu\'ery-Odelin, A.~Ruschhaupt, A.~Kiely, E.~Torrontegui,
  S.~Mart\'{\i}nez-Garaot, J.~G. Muga,
  \href{https://link.aps.org/doi/10.1103/RevModPhys.91.045001}{Shortcuts to
  adiabaticity: Concepts, methods, and applications}, Rev. Mod. Phys. 91 (2019)
  045001.
\newblock \href {https://doi.org/10.1103/RevModPhys.91.045001}
  {\path{doi:10.1103/RevModPhys.91.045001}}.
\newline\urlprefix\url{https://link.aps.org/doi/10.1103/RevModPhys.91.045001}

\bibitem{DRAG_2009}
F.~Motzoi, J.~M. Gambetta, P.~Rebentrost, F.~K. Wilhelm,
  \href{https://link.aps.org/doi/10.1103/PhysRevLett.103.110501}{Simple pulses
  for elimination of leakage in weakly nonlinear qubits}, Phys. Rev. Lett. 103
  (2009) 110501.
\newblock \href {https://doi.org/10.1103/PhysRevLett.103.110501}
  {\path{doi:10.1103/PhysRevLett.103.110501}}.
\newline\urlprefix\url{https://link.aps.org/doi/10.1103/PhysRevLett.103.110501}

\bibitem{DD_1950}
E.~L. Hahn, \href{https://link.aps.org/doi/10.1103/PhysRev.80.580}{Spin
  echoes}, Phys. Rev. 80 (1950) 580--594.
\newblock \href {https://doi.org/10.1103/PhysRev.80.580}
  {\path{doi:10.1103/PhysRev.80.580}}.
\newline\urlprefix\url{https://link.aps.org/doi/10.1103/PhysRev.80.580}

\bibitem{DD_1954}
H.~Y. Carr, E.~M. Purcell,
  \href{https://link.aps.org/doi/10.1103/PhysRev.94.630}{Effects of diffusion
  on free precession in nuclear magnetic resonance experiments}, Phys. Rev. 94
  (1954) 630--638.
\newblock \href {https://doi.org/10.1103/PhysRev.94.630}
  {\path{doi:10.1103/PhysRev.94.630}}.
\newline\urlprefix\url{https://link.aps.org/doi/10.1103/PhysRev.94.630}

\bibitem{DD_1999}
L.~Viola, E.~Knill, S.~Lloyd,
  \href{https://link.aps.org/doi/10.1103/PhysRevLett.82.2417}{Dynamical
  decoupling of open quantum systems}, Phys. Rev. Lett. 82 (1999) 2417--2421.
\newblock \href {https://doi.org/10.1103/PhysRevLett.82.2417}
  {\path{doi:10.1103/PhysRevLett.82.2417}}.
\newline\urlprefix\url{https://link.aps.org/doi/10.1103/PhysRevLett.82.2417}

\bibitem{DD_2003}
L.~Viola, E.~Knill,
  \href{https://link.aps.org/doi/10.1103/PhysRevLett.90.037901}{Robust
  dynamical decoupling of quantum systems with bounded controls}, Phys. Rev.
  Lett. 90 (2003) 037901.
\newblock \href {https://doi.org/10.1103/PhysRevLett.90.037901}
  {\path{doi:10.1103/PhysRevLett.90.037901}}.
\newline\urlprefix\url{https://link.aps.org/doi/10.1103/PhysRevLett.90.037901}

\bibitem{DD_2005}
K.~Khodjasteh, D.~A. Lidar,
  \href{https://link.aps.org/doi/10.1103/PhysRevLett.95.180501}{Fault-tolerant
  quantum dynamical decoupling}, Phys. Rev. Lett. 95 (2005) 180501.
\newblock \href {https://doi.org/10.1103/PhysRevLett.95.180501}
  {\path{doi:10.1103/PhysRevLett.95.180501}}.
\newline\urlprefix\url{https://link.aps.org/doi/10.1103/PhysRevLett.95.180501}

\bibitem{DD_2009}
M.~J. Biercuk, H.~Uys, A.~P. VanDevender, N.~Shiga, W.~M. Itano, J.~J.
  Bollinger, \href{https://doi.org/10.1038/nature07951}{Optimized dynamical
  decoupling in a model quantum memory}, Nature 458~(7241) (2009) 996--1000.
\newblock \href {https://doi.org/10.1038/nature07951}
  {\path{doi:10.1038/nature07951}}.
\newline\urlprefix\url{https://doi.org/10.1038/nature07951}

\bibitem{DD_2010}
J.~R. West, D.~A. Lidar, B.~H. Fong, M.~F. Gyure,
  \href{https://link.aps.org/doi/10.1103/PhysRevLett.105.230503}{High fidelity
  quantum gates via dynamical decoupling}, Phys. Rev. Lett. 105 (2010) 230503.
\newblock \href {https://doi.org/10.1103/PhysRevLett.105.230503}
  {\path{doi:10.1103/PhysRevLett.105.230503}}.
\newline\urlprefix\url{https://link.aps.org/doi/10.1103/PhysRevLett.105.230503}

\bibitem{Degen_2017}
C.~L. Degen, F.~Reinhard, P.~Cappellaro,
  \href{https://link.aps.org/doi/10.1103/RevModPhys.89.035002}{Quantum
  sensing}, Rev. Mod. Phys. 89 (2017) 035002.
\newblock \href {https://doi.org/10.1103/RevModPhys.89.035002}
  {\path{doi:10.1103/RevModPhys.89.035002}}.
\newline\urlprefix\url{https://link.aps.org/doi/10.1103/RevModPhys.89.035002}

\bibitem{muller2018noise}
M.~M. M{\"u}ller, S.~Gherardini, F.~Caruso, Noise-robust quantum sensing via
  optimal multi-probe spectroscopy, Scientific reports 8~(1) (2018) 1--17.

\bibitem{Petruhanov_2022}
V.~N. Petruhanov, A.~N. Pechen,
  \href{https://doi.org/10.1142/S0217751X22430175}{Optimal control for state
  preparation in two-qubit open quantum systems driven by coherent and
  incoherent controls via grape approach}, International Journal of Modern
  Physics A 37~(20n21) (2022) 2243017.
\newblock \href
  {http://arxiv.org/abs/https://doi.org/10.1142/S0217751X22430175}
  {\path{arXiv:https://doi.org/10.1142/S0217751X22430175}}, \href
  {https://doi.org/10.1142/S0217751X22430175}
  {\path{doi:10.1142/S0217751X22430175}}.
\newline\urlprefix\url{https://doi.org/10.1142/S0217751X22430175}

\bibitem{Preti_2022}
F.~Preti, T.~Calarco, J.~M. Torres, J.~Z. Bernád,
  \href{https://arxiv.org/abs/2205.12091}{Optimal two-qubit gates in recurrence
  protocols of entanglement purification} (2022).
\newblock \href {https://doi.org/10.48550/ARXIV.2205.12091}
  {\path{doi:10.48550/ARXIV.2205.12091}}.
\newline\urlprefix\url{https://arxiv.org/abs/2205.12091}

\bibitem{tomlinsonFourierSynthesizedExcitation1973}
B.~L. Tomlinson, H.~D.~W. Hill, Fourier synthesized excitation of nuclear
  magnetic resonance with application to homonuclear decoupling and solvent
  line suppression, The Journal of Chemical Physics 59~(4) (1973) 1775--1784.
\newblock \href {https://doi.org/10.1063/1.1680263}
  {\path{doi:10.1063/1.1680263}}.

\bibitem{Tannor1992}
D.~J. Tannor, V.~Kazakov, V.~Orlov, Control of photochemical branching:
  {{Novel}} procedures for finding optimal pulses and global upper bounds, in:
  J.~Broeckhove, L.~Lathouwers (Eds.), Time-Dependent Quantum Molecular
  Dynamics, {Springer US}, {Boston, MA}, 1992, pp. 347--360.
\newblock \href {https://doi.org/10.1007/978-1-4899-2326-4_24}
  {\path{doi:10.1007/978-1-4899-2326-4_24}}.

\bibitem{goerzKrotovPythonImplementation2019}
M.~Goerz, D.~Basilewitsch, F.~{Gago-Encinas}, M.~G. Krauss, K.~P. Horn, D.~M.
  Reich, C.~Koch, Krotov: {{A Python}} implementation of {{Krotov}}'s method
  for quantum optimal control, SciPost Physics 7~(6) (2019) 080.
\newblock \href {https://doi.org/10.21468/SciPostPhys.7.6.080}
  {\path{doi:10.21468/SciPostPhys.7.6.080}}.

\bibitem{reichMonotonicallyConvergentOptimization2012}
D.~M. Reich, M.~Ndong, C.~P. Koch,
  \href{https://doi.org/10.1063/1.3691827}{Monotonically convergent
  optimization in quantum control using {{Krotov}}'s method}, J. Chem. Phys.
  136~(10) (2012) 104103.
\newblock \href {https://doi.org/10.1063/1.3691827}
  {\path{doi:10.1063/1.3691827}}.
\newline\urlprefix\url{https://doi.org/10.1063/1.3691827}

\bibitem{khaneja2005optimal}
N.~Khaneja, T.~Reiss, C.~Kehlet, T.~Schulte-Herbrüggen, S.~J. Glaser,
  \href{https://www.sciencedirect.com/science/article/pii/S1090780704003696}{Optimal
  control of coupled spin dynamics: design of {NMR} pulse sequences by gradient
  ascent algorithms}, Journal of Magnetic Resonance 172~(2) (2005) 296--305.
\newblock \href {https://doi.org/https://doi.org/10.1016/j.jmr.2004.11.004}
  {\path{doi:https://doi.org/10.1016/j.jmr.2004.11.004}}.
\newline\urlprefix\url{https://www.sciencedirect.com/science/article/pii/S1090780704003696}

\bibitem{Machnes2011}
S.~Machnes, U.~Sander, S.~J. Glaser, P.~de~Fouqui\`eres, A.~Gruslys,
  S.~Schirmer, T.~Schulte-Herbr\"uggen,
  \href{https://link.aps.org/doi/10.1103/PhysRevA.84.022305}{Comparing,
  optimizing, and benchmarking quantum-control algorithms in a unifying
  programming framework}, Phys. Rev. A 84 (2011) 022305.
\newblock \href {https://doi.org/10.1103/PhysRevA.84.022305}
  {\path{doi:10.1103/PhysRevA.84.022305}}.
\newline\urlprefix\url{https://link.aps.org/doi/10.1103/PhysRevA.84.022305}

\bibitem{Doria_2011}
P.~Doria, T.~Calarco, S.~Montangero,
  \href{https://link.aps.org/doi/10.1103/PhysRevLett.106.190501}{Optimal
  control technique for many-body quantum dynamics}, Phys. Rev. Lett. 106
  (2011) 190501.
\newblock \href {https://doi.org/10.1103/PhysRevLett.106.190501}
  {\path{doi:10.1103/PhysRevLett.106.190501}}.
\newline\urlprefix\url{https://link.aps.org/doi/10.1103/PhysRevLett.106.190501}

\bibitem{canevaChoppedRandombasisQuantum2011a}
T.~Caneva, T.~Calarco, S.~Montangero,
  \href{https://link.aps.org/doi/10.1103/PhysRevA.84.022326}{Chopped
  random-basis quantum optimization}, Phys. Rev. A 84 (2011) 022326.
\newblock \href {https://doi.org/10.1103/PhysRevA.84.022326}
  {\path{doi:10.1103/PhysRevA.84.022326}}.
\newline\urlprefix\url{https://link.aps.org/doi/10.1103/PhysRevA.84.022326}

\bibitem{rachDressingChoppedrandombasisOptimization2015a}
N.~Rach, M.~M. M\"uller, T.~Calarco, S.~Montangero,
  \href{https://link.aps.org/doi/10.1103/PhysRevA.92.062343}{Dressing the
  chopped-random-basis optimization: A bandwidth-limited access to the
  trap-free landscape}, Phys. Rev. A 92 (2015) 062343.
\newblock \href {https://doi.org/10.1103/PhysRevA.92.062343}
  {\path{doi:10.1103/PhysRevA.92.062343}}.
\newline\urlprefix\url{https://link.aps.org/doi/10.1103/PhysRevA.92.062343}

\bibitem{johansson2012qutip}
J.~R. Johansson, P.~D. Nation, F.~Nori, Qutip: An open-source python framework
  for the dynamics of open quantum systems, Computer Physics Communications
  183~(8) (2012) 1760--1772.

\bibitem{teskeQoptExperimentorientedQubit2022}
J.~D. Teske, P.~Cerfontaine, H.~Bluhm,
  \href{https://link.aps.org/doi/10.1103/PhysRevApplied.17.034036}{Qopt: {{An}}
  experiment-oriented {{Qubit Simulation}} and {{Quantum Optimal Control
  Package}}}, Phys. Rev. Applied 17 (2022) 034036.
\newblock \href {https://doi.org/10.1103/PhysRevApplied.17.034036}
  {\path{doi:10.1103/PhysRevApplied.17.034036}}.
\newline\urlprefix\url{https://link.aps.org/doi/10.1103/PhysRevApplied.17.034036}

\bibitem{HOGBEN2011179}
H.~J. Hogben, M.~Krzystyniak, G.~T.~P. Charnock, P.~J. Hore, I.~Kuprov,
  \href{https://www.sciencedirect.com/science/article/pii/S1090780710003575}{{Spinach
  – A software library for simulation of spin dynamics in large spin
  systems}}, Journal of Magnetic Resonance 208~(2) (2011) 179--194.
\newblock \href {https://doi.org/https://doi.org/10.1016/j.jmr.2010.11.008}
  {\path{doi:https://doi.org/10.1016/j.jmr.2010.11.008}}.
\newline\urlprefix\url{https://www.sciencedirect.com/science/article/pii/S1090780710003575}

\bibitem{Wittler_2021}
N.~Wittler, F.~Roy, K.~Pack, M.~Werninghaus, A.~S. Roy, D.~J. Egger, S.~Filipp,
  F.~K. Wilhelm, S.~Machnes,
  \href{https://link.aps.org/doi/10.1103/PhysRevApplied.15.034080}{Integrated
  tool set for control, calibration, and characterization of quantum devices
  applied to superconducting qubits}, Phys. Rev. Applied 15 (2021) 034080.
\newblock \href {https://doi.org/10.1103/PhysRevApplied.15.034080}
  {\path{doi:10.1103/PhysRevApplied.15.034080}}.
\newline\urlprefix\url{https://link.aps.org/doi/10.1103/PhysRevApplied.15.034080}

\bibitem{Binder2017}
J.~M. Binder, A.~Stark, N.~Tomek, J.~Scheuer, F.~Frank, K.~D. Jahnke,
  C.~M{\"u}ller, S.~Schmitt, M.~H. Metsch, T.~Unden, T.~Gehring, A.~Huck, U.~L.
  Andersen, L.~J. Rogers, F.~Jelezko, Qudi: A modular python suite for
  experiment control and data processing, SoftwareX 6 (2017) 85 -- 90.
\newblock \href {https://doi.org/https://doi.org/10.1016/j.softx.2017.02.001}
  {\path{doi:https://doi.org/10.1016/j.softx.2017.02.001}}.

\bibitem{scheuer2014precise}
J.~Scheuer, X.~Kong, R.~S. Said, J.~Chen, A.~Kurz, L.~Marseglia, J.~Du, P.~R.
  Hemmer, S.~Montangero, T.~Calarco, et~al., Precise qubit control beyond the
  rotating wave approximation, New Journal of Physics 16~(9) (2014) 093022.

\bibitem{waldherr2014quantum}
G.~Waldherr, Y.~Wang, S.~Zaiser, M.~Jamali, T.~Schulte-Herbr{\"u}ggen, H.~Abe,
  T.~Ohshima, J.~Isoya, J.~Du, P.~Neumann, et~al., Quantum error correction in
  a solid-state hybrid spin register, Nature 506~(7487) (2014) 204--207.

\bibitem{dolde2014high}
F.~Dolde, V.~Bergholm, Y.~Wang, I.~Jakobi, B.~Naydenov, S.~Pezzagna, J.~Meijer,
  F.~Jelezko, P.~Neumann, T.~Schulte-Herbr{\"u}ggen, et~al., High-fidelity spin
  entanglement using optimal control, Nature communications 5~(1) (2014) 1--9.

\bibitem{unden2016quantum}
T.~Unden, P.~Balasubramanian, D.~Louzon, Y.~Vinkler, M.~B. Plenio, M.~Markham,
  D.~Twitchen, A.~Stacey, I.~Lovchinsky, A.~O. Sushkov, et~al., Quantum
  metrology enhanced by repetitive quantum error correction, Physical review
  letters 116~(23) (2016) 230502.

\bibitem{Frank2017}
F.~Frank, T.~Unden, J.~Zoller, R.~S. Said, T.~Calarco, S.~Montangero,
  B.~Naydenov, F.~Jelezko, Autonomous calibration of single spin qubit
  operations, npj Quantum Information 3~(1) (2017) 48.
\newblock \href {https://doi.org/10.1038/s41534-017-0049-8}
  {\path{doi:10.1038/s41534-017-0049-8}}.

\bibitem{schmitt2017submillihertz}
S.~Schmitt, T.~Gefen, F.~M. St{\"u}rner, T.~Unden, G.~Wolff, C.~M{\"u}ller,
  J.~Scheuer, B.~Naydenov, M.~Markham, S.~Pezzagna, et~al., Submillihertz
  magnetic spectroscopy performed with a nanoscale quantum sensor, Science
  356~(6340) (2017) 832--837.

\bibitem{poggiali2018optimal}
F.~Poggiali, P.~Cappellaro, N.~Fabbri, Optimal control for one-qubit quantum
  sensing, Physical Review X 8~(2) (2018) 021059.

\bibitem{Oshnik_2022}
N.~Oshnik, P.~Rembold, T.~Calarco, S.~Montangero, E.~Neu, M.~M. M\"uller,
  \href{https://link.aps.org/doi/10.1103/PhysRevA.106.013107}{Robust
  magnetometry with single nitrogen-vacancy centers via two-step optimization},
  Phys. Rev. A 106 (2022) 013107.
\newblock \href {https://doi.org/10.1103/PhysRevA.106.013107}
  {\path{doi:10.1103/PhysRevA.106.013107}}.
\newline\urlprefix\url{https://link.aps.org/doi/10.1103/PhysRevA.106.013107}

\bibitem{Conolly_1986}
S.~Conolly, D.~Nishimura, A.~Macovski, Optimal control solutions to the
  magnetic resonance selective excitation problem, IEEE Transactions on Medical
  Imaging 5~(2) (1986) 106--115.
\newblock \href {https://doi.org/10.1109/TMI.1986.4307754}
  {\path{doi:10.1109/TMI.1986.4307754}}.

\bibitem{Peirce_1988}
A.~P. Peirce, M.~A. Dahleh, H.~Rabitz,
  \href{https://link.aps.org/doi/10.1103/PhysRevA.37.4950}{Optimal control of
  quantum-mechanical systems: Existence, numerical approximation, and
  applications}, Phys. Rev. A 37 (1988) 4950--4964.
\newblock \href {https://doi.org/10.1103/PhysRevA.37.4950}
  {\path{doi:10.1103/PhysRevA.37.4950}}.
\newline\urlprefix\url{https://link.aps.org/doi/10.1103/PhysRevA.37.4950}

\bibitem{McDonald_1991}
S.~McDonald, W.~S. Warren,
  \href{https://onlinelibrary.wiley.com/doi/abs/10.1002/cmr.1820030202}{Uses of
  shaped pulses in nmr: A primer}, Concepts in Magnetic Resonance 3~(2) (1991)
  55--81.
\newblock \href
  {http://arxiv.org/abs/https://onlinelibrary.wiley.com/doi/pdf/10.1002/cmr.1820030202}
  {\path{arXiv:https://onlinelibrary.wiley.com/doi/pdf/10.1002/cmr.1820030202}},
  \href {https://doi.org/https://doi.org/10.1002/cmr.1820030202}
  {\path{doi:https://doi.org/10.1002/cmr.1820030202}}.
\newline\urlprefix\url{https://onlinelibrary.wiley.com/doi/abs/10.1002/cmr.1820030202}

\bibitem{Casanova_2018}
J.~Casanova, Z.-Y. Wang, I.~Schwartz, M.~B. Plenio,
  \href{https://link.aps.org/doi/10.1103/PhysRevApplied.10.044072}{Shaped
  pulses for energy-efficient high-field nmr at the nanoscale}, Phys. Rev.
  Applied 10 (2018) 044072.
\newblock \href {https://doi.org/10.1103/PhysRevApplied.10.044072}
  {\path{doi:10.1103/PhysRevApplied.10.044072}}.
\newline\urlprefix\url{https://link.aps.org/doi/10.1103/PhysRevApplied.10.044072}

\bibitem{Vetter_2021}
P.~J. Vetter, A.~Marshall, G.~T. Genov, T.~F. Weiss, N.~Striegler, E.~F.
  Gro\ss{}mann, S.~Oviedo-Casado, J.~Cerrillo, J.~Prior, P.~Neumann,
  F.~Jelezko,
  \href{https://link.aps.org/doi/10.1103/PhysRevApplied.17.044028}{Zero- and
  low-field sensing with nitrogen-vacancy centers}, Phys. Rev. Applied 17
  (2022) 044028.
\newblock \href {https://doi.org/10.1103/PhysRevApplied.17.044028}
  {\path{doi:10.1103/PhysRevApplied.17.044028}}.
\newline\urlprefix\url{https://link.aps.org/doi/10.1103/PhysRevApplied.17.044028}

\bibitem{muller2015phonon}
M.~M. M{\"u}ller, U.~G. Poschinger, T.~Calarco, S.~Montangero,
  F.~Schmidt-Kaler, Phonon-to-spin mapping in a system of a trapped ion via
  optimal control, Physical Review A 92~(5) (2015) 053423.

\bibitem{furst2014controlling}
H.~F{\"u}rst, M.~H. Goerz, U.~Poschinger, M.~Murphy, S.~Montangero, T.~Calarco,
  F.~Schmidt-Kaler, K.~Singer, C.~P. Koch, Controlling the transport of an ion:
  classical and quantum mechanical solutions, New Journal of Physics 16~(7)
  (2014) 075007.

\bibitem{pichler2016noise}
T.~Pichler, T.~Caneva, S.~Montangero, M.~D. Lukin, T.~Calarco, Noise-resistant
  optimal spin squeezing via quantum control, Physical Review A 93~(1) (2016)
  013851.

\bibitem{monz201114}
T.~Monz, P.~Schindler, J.~T. Barreiro, M.~Chwalla, D.~Nigg, W.~A. Coish,
  M.~Harlander, W.~H{\"a}nsel, M.~Hennrich, R.~Blatt, 14-qubit entanglement:
  Creation and coherence, Physical Review Letters 106~(13) (2011) 130506.

\bibitem{walther2012controlling}
A.~Walther, F.~Ziesel, T.~Ruster, S.~T. Dawkins, K.~Ott, M.~Hettrich,
  K.~Singer, F.~Schmidt-Kaler, U.~Poschinger, Controlling fast transport of
  cold trapped ions, Physical review letters 109~(8) (2012) 080501.

\bibitem{casanova2011quantum}
J.~Casanova, L.~Lamata, I.~Egusquiza, R.~Gerritsma, C.~F. Roos, J.~J.
  Garc{\'\i}a-Ripoll, E.~Solano, Quantum simulation of quantum field theories
  in trapped ions, Physical review letters 107~(26) (2011) 260501.

\bibitem{singer2010colloquium}
K.~Singer, U.~Poschinger, M.~Murphy, P.~Ivanov, F.~Ziesel, T.~Calarco,
  F.~Schmidt-Kaler, Colloquium: Trapped ions as quantum bits: Essential
  numerical tools, Reviews of modern physics 82~(3) (2010) 2609.

\bibitem{zhang2018noon}
J.~Zhang, M.~Um, D.~Lv, J.-N. Zhang, L.-M. Duan, K.~Kim, Noon states of nine
  quantized vibrations in two radial modes of a trapped ion, Physical review
  letters 121~(16) (2018) 160502.

\bibitem{leibfried2003quantum}
D.~Leibfried, R.~Blatt, C.~Monroe, D.~Wineland, Quantum dynamics of single
  trapped ions, Reviews of Modern Physics 75~(1) (2003) 281.

\bibitem{rosi2013fast}
S.~Rosi, A.~Bernard, N.~Fabbri, L.~Fallani, C.~Fort, M.~Inguscio, T.~Calarco,
  S.~Montangero, Fast closed-loop optimal control of ultracold atoms in an
  optical lattice, Physical Review A 88~(2) (2013) 021601.

\bibitem{van2014interferometry}
S.~van Frank, A.~Negretti, T.~Berrada, R.~B{\"u}cker, S.~Montangero, J.-F.
  Schaff, T.~Schumm, T.~Calarco, J.~Schmiedmayer, Interferometry with
  non-classical motional states of a bose--einstein condensate, Nature
  communications 5~(1) (2014) 1--6.

\bibitem{van2016optimal}
S.~van Frank, M.~Bonneau, J.~Schmiedmayer, S.~Hild, C.~Gross, M.~Cheneau,
  I.~Bloch, T.~Pichler, A.~Negretti, T.~Calarco, et~al., Optimal control of
  complex atomic quantum systems, Scientific reports 6~(1) (2016) 1--12.

\bibitem{brouzos2015quantum}
I.~Brouzos, A.~I. Streltsov, A.~Negretti, R.~S. Said, T.~Caneva, S.~Montangero,
  T.~Calarco, Quantum speed limit and optimal control of many-boson dynamics,
  Physical Review A 92~(6) (2015) 062110.

\bibitem{sorensen2018quantum}
J.~S{\o}rensen, M.~Aranburu, T.~Heinzel, J.~Sherson, Quantum optimal control in
  a chopped basis: Applications in control of bose-einstein condensates,
  Physical Review A 98~(2) (2018) 022119.

\bibitem{heck2018remote}
R.~Heck, O.~Vuculescu, J.~J. S{\o}rensen, J.~Zoller, M.~G. Andreasen, M.~G.
  Bason, P.~Ejlertsen, O.~El{\'\i}asson, P.~Haikka, J.~S. Laustsen, et~al.,
  Remote optimization of an ultracold atoms experiment by experts and citizen
  scientists, Proceedings of the National Academy of Sciences 115~(48) (2018)
  E11231--E11237.

\bibitem{omran2019generation}
A.~Omran, H.~Levine, A.~Keesling, G.~Semeghini, T.~T. Wang, S.~Ebadi,
  H.~Bernien, A.~S. Zibrov, H.~Pichler, S.~Choi, et~al., Generation and
  manipulation of schr{\"o}dinger cat states in rydberg atom arrays, Science
  365~(6453) (2019) 570--574.

\bibitem{Mastroserio2022}
I.~Mastroserio, S.~Gherardini, C.~Lovecchio, T.~Calarco, S.~Montangero, F.~S.
  Cataliotti, F.~Caruso, \href{https://arxiv.org/abs/2206.02746}{Experimental
  realization of optimal time-reversal on an atom chip for quantum undo
  operations} (2022).
\newblock \href {https://doi.org/10.48550/ARXIV.2206.02746}
  {\path{doi:10.48550/ARXIV.2206.02746}}.
\newline\urlprefix\url{https://arxiv.org/abs/2206.02746}

\bibitem{watts2015optimizing}
P.~Watts, J.~Vala, M.~M. M{\"u}ller, T.~Calarco, K.~B. Whaley, D.~M. Reich,
  M.~H. Goerz, C.~P. Koch, Optimizing for an arbitrary perfect entangler. i.
  functionals, Physical Review A 91~(6) (2015) 062306.

\bibitem{goerz2015optimizing}
M.~H. Goerz, G.~Gualdi, D.~M. Reich, C.~P. Koch, F.~Motzoi, K.~B. Whaley,
  J.~Vala, M.~M. M{\"u}ller, S.~Montangero, T.~Calarco, Optimizing for an
  arbitrary perfect entangler. ii. application, Physical Review A 91~(6) (2015)
  062307.

\bibitem{hoeb2017amplification}
F.~Hoeb, F.~Angaroni, J.~Zoller, T.~Calarco, G.~Strini, S.~Montangero,
  G.~Benenti, Amplification of the parametric dynamical casimir effect via
  optimal control, Physical Review A 96~(3) (2017) 033851.

\bibitem{Heeres2017}
R.~W. Heeres, P.~Reinhold, N.~Ofek, L.~Frunzio, L.~Jiang, M.~H. Devoret, R.~J.
  Schoelkopf, \href{https://doi.org/10.1038/s41467-017-00045-1}{Implementing a
  universal gate set on a logical qubit encoded in an oscillator}, Nature
  Communications 8~(1) (2017) 94.
\newblock \href {https://doi.org/10.1038/s41467-017-00045-1}
  {\path{doi:10.1038/s41467-017-00045-1}}.
\newline\urlprefix\url{https://doi.org/10.1038/s41467-017-00045-1}

\bibitem{Theis_2018}
L.~S. Theis, F.~Motzoi, S.~Machnes, F.~K. Wilhelm,
  \href{https://doi.org/10.1209/0295-5075/123/60001}{Counteracting systems of
  diabaticities using {DRAG} controls: The status after 10 years}, {EPL}
  (Europhysics Letters) 123~(6) (2018) 60001.
\newblock \href {https://doi.org/10.1209/0295-5075/123/60001}
  {\path{doi:10.1209/0295-5075/123/60001}}.
\newline\urlprefix\url{https://doi.org/10.1209/0295-5075/123/60001}

\bibitem{Heck2018}
R.~Heck, O.~Vuculescu, J.~J. S{\o}rensen, J.~Zoller, M.~G. Andreasen, M.~G.
  Bason, P.~Ejlertsen, O.~El{\'\i}asson, P.~Haikka, J.~S. Laustsen, L.~L.
  Nielsen, A.~Mao, R.~M{\"u}ller, M.~Napolitano, M.~K. Pedersen, A.~R. Thorsen,
  C.~Bergenholtz, T.~Calarco, S.~Montangero, J.~F. Sherson, Remote optimization
  of an ultracold atoms experiment by experts and citizen scientists,
  Proceedings of the National Academy of Sciences 115~(48) (2018)
  E11231--E11237.
\newblock \href {https://doi.org/10.1073/pnas.1716869115}
  {\path{doi:10.1073/pnas.1716869115}}.

\bibitem{Angaroni_2019}
F.~Angaroni, A.~Graudenzi, M.~Rossignolo, D.~Maspero, T.~Calarco, R.~Piazza,
  S.~Montangero, M.~Antoniotti,
  \href{https://www.biorxiv.org/content/early/2019/06/07/662858}{Personalized
  therapy design for liquid tumors via optimal control theory}, bioRxiv (2019).
\newblock \href
  {http://arxiv.org/abs/https://www.biorxiv.org/content/early/2019/06/07/662858.full.pdf}
  {\path{arXiv:https://www.biorxiv.org/content/early/2019/06/07/662858.full.pdf}},
  \href {https://doi.org/10.1101/662858} {\path{doi:10.1101/662858}}.
\newline\urlprefix\url{https://www.biorxiv.org/content/early/2019/06/07/662858}

\bibitem{macroscopic_hyperpol_2022}
A.~Marshall, T.~Reisser, P.~Rembold, C.~M\"uller, J.~Scheuer, M.~Gierse,
  T.~Eichhorn, J.~M. Steiner, P.~Hautle, T.~Calarco, F.~Jelezko, M.~B. Plenio,
  S.~Montangero, I.~Schwartz, M.~M. M\"uller, P.~Neumann,
  \href{https://link.aps.org/doi/10.1103/PhysRevResearch.4.043179}{Macroscopic
  hyperpolarization enhanced with quantum optimal control}, Phys. Rev. Res. 4
  (2022) 043179.
\newblock \href {https://doi.org/10.1103/PhysRevResearch.4.043179}
  {\path{doi:10.1103/PhysRevResearch.4.043179}}.
\newline\urlprefix\url{https://link.aps.org/doi/10.1103/PhysRevResearch.4.043179}

\bibitem{Pagano2022}
A.~Pagano, S.~Weber, D.~Jaschke, T.~Pfau, F.~Meinert, S.~Montangero, H.~P.
  B\"uchler,
  \href{https://link.aps.org/doi/10.1103/PhysRevResearch.4.033019}{Error
  budgeting for a controlled-phase gate with strontium-88 rydberg atoms}, Phys.
  Rev. Research 4 (2022) 033019.
\newblock \href {https://doi.org/10.1103/PhysRevResearch.4.033019}
  {\path{doi:10.1103/PhysRevResearch.4.033019}}.
\newline\urlprefix\url{https://link.aps.org/doi/10.1103/PhysRevResearch.4.033019}

\bibitem{Marquardt_2008}
F.~Marquardt, A.~Püttmann, \href{https://arxiv.org/abs/0809.4403}{Introduction
  to dissipation and decoherence in quantum systems} (2008).
\newblock \href {https://doi.org/10.48550/ARXIV.0809.4403}
  {\path{doi:10.48550/ARXIV.0809.4403}}.
\newline\urlprefix\url{https://arxiv.org/abs/0809.4403}

\bibitem{Nelder1965}
J.~A. Nelder, R.~Mead, {A Simplex Method for Function Minimization}, The
  Computer Journal 7~(4) (1965) 308--313.
\newblock \href
  {http://arxiv.org/abs/http://oup.prod.sis.lan/comjnl/article-pdf/7/4/308/1013182/7-4-308.pdf}
  {\path{arXiv:http://oup.prod.sis.lan/comjnl/article-pdf/7/4/308/1013182/7-4-308.pdf}},
  \href {https://doi.org/10.1093/comjnl/7.4.308}
  {\path{doi:10.1093/comjnl/7.4.308}}.

\bibitem{gao2012implementing}
F.~Gao, L.~Han, Implementing the nelder-mead simplex algorithm with adaptive
  parameters, Computational Optimization and Applications 51~(1) (2012)
  259--277.

\bibitem{Powell1964}
M.~J.~D. Powell, {An efficient method for finding the minimum of a function of
  several variables without calculating derivatives}, The Computer Journal
  7~(2) (1964) 155--162.
\newblock \href
  {http://arxiv.org/abs/http://oup.prod.sis.lan/comjnl/article-pdf/7/2/155/959784/070155.pdf}
  {\path{arXiv:http://oup.prod.sis.lan/comjnl/article-pdf/7/2/155/959784/070155.pdf}},
  \href {https://doi.org/10.1093/comjnl/7.2.155}
  {\path{doi:10.1093/comjnl/7.2.155}}.

\bibitem{Hansen2003}
N.~Hansen, S.~D. M{\"u}ller, P.~Koumoutsakos, Reducing the time complexity of
  the derandomized evolution strategy with covariance matrix adaptation
  (cma-es), Evolutionary Computation 11~(1) (2003) 1--18.
\newblock \href
  {http://arxiv.org/abs/https://doi.org/10.1162/106365603321828970}
  {\path{arXiv:https://doi.org/10.1162/106365603321828970}}, \href
  {https://doi.org/10.1162/106365603321828970}
  {\path{doi:10.1162/106365603321828970}}.

\bibitem{Liu1989}
D.~C. Liu, J.~Nocedal, \href{https://doi.org/10.1007/BF01589116}{On the limited
  memory bfgs method for large scale optimization}, Mathematical Programming
  45~(1) (1989) 503--528.
\newblock \href {https://doi.org/10.1007/BF01589116}
  {\path{doi:10.1007/BF01589116}}.
\newline\urlprefix\url{https://doi.org/10.1007/BF01589116}

\bibitem{jupyter_notebooks}
Quantum-OCS,
  \href{https://github.com/Quantum-OCS/QuOCS-jupyternotebooks}{Jupyter
  notebooks for {QuOCS}}, {Accessed}: 2022-12-07.
\newline\urlprefix\url{https://github.com/Quantum-OCS/QuOCS-jupyternotebooks}

\bibitem{Felix_deep_learning_2022}
M.~Dalgaard, F.~Motzoi, J.~Sherson, Predicting quantum dynamical cost
  landscapes with deep learning, Physical Review A 105~(1) (Jan 2022).
\newblock \href {https://doi.org/10.1103/physreva.105.012402}
  {\path{doi:10.1103/physreva.105.012402}}.

\bibitem{SciPy}
{SciPy Documentation}, {Settings for L-BFGS-B},
  \url{https://docs.scipy.org/doc/scipy/reference/optimize.minimize-lbfgsb.html},
  accessed: 2022\_08\_19.

\bibitem{Soerensen_2018}
J.~J. W.~H. S\o{}rensen, M.~O. Aranburu, T.~Heinzel, J.~F. Sherson,
  \href{https://link.aps.org/doi/10.1103/PhysRevA.98.022119}{Quantum optimal
  control in a chopped basis: Applications in control of bose-einstein
  condensates}, Phys. Rev. A 98 (2018) 022119.
\newblock \href {https://doi.org/10.1103/PhysRevA.98.022119}
  {\path{doi:10.1103/PhysRevA.98.022119}}.
\newline\urlprefix\url{https://link.aps.org/doi/10.1103/PhysRevA.98.022119}

\bibitem{Machnes_2018}
S.~Machnes, E.~Ass\'emat, D.~Tannor, F.~K. Wilhelm,
  \href{https://link.aps.org/doi/10.1103/PhysRevLett.120.150401}{Tunable,
  flexible, and efficient optimization of control pulses for practical qubits},
  Phys. Rev. Lett. 120 (2018) 150401.
\newblock \href {https://doi.org/10.1103/PhysRevLett.120.150401}
  {\path{doi:10.1103/PhysRevLett.120.150401}}.
\newline\urlprefix\url{https://link.aps.org/doi/10.1103/PhysRevLett.120.150401}

\bibitem{Motzoi_2011}
F.~Motzoi, J.~M. Gambetta, S.~T. Merkel, F.~K. Wilhelm,
  \href{https://link.aps.org/doi/10.1103/PhysRevA.84.022307}{Optimal control
  methods for rapidly time-varying hamiltonians}, Phys. Rev. A 84 (2011)
  022307.
\newblock \href {https://doi.org/10.1103/PhysRevA.84.022307}
  {\path{doi:10.1103/PhysRevA.84.022307}}.
\newline\urlprefix\url{https://link.aps.org/doi/10.1103/PhysRevA.84.022307}

\bibitem{DOHERTY20131}
M.~W. Doherty, N.~B. Manson, P.~Delaney, F.~Jelezko, J.~Wrachtrup, L.~C.
  Hollenberg,
  \href{https://www.sciencedirect.com/science/article/pii/S0370157313000562}{The
  nitrogen-vacancy colour centre in diamond}, Physics Reports 528~(1) (2013)
  1--45, the nitrogen-vacancy colour centre in diamond.
\newblock \href {https://doi.org/https://doi.org/10.1016/j.physrep.2013.02.001}
  {\path{doi:https://doi.org/10.1016/j.physrep.2013.02.001}}.
\newline\urlprefix\url{https://www.sciencedirect.com/science/article/pii/S0370157313000562}

\bibitem{PhysRevLett.102.057403}
V.~Jacques, P.~Neumann, J.~Beck, M.~Markham, D.~Twitchen, J.~Meijer, F.~Kaiser,
  G.~Balasubramanian, F.~Jelezko, J.~Wrachtrup,
  \href{https://link.aps.org/doi/10.1103/PhysRevLett.102.057403}{Dynamic
  polarization of single nuclear spins by optical pumping of nitrogen-vacancy
  color centers in diamond at room temperature}, Phys. Rev. Lett. 102 (2009)
  057403.
\newblock \href {https://doi.org/10.1103/PhysRevLett.102.057403}
  {\path{doi:10.1103/PhysRevLett.102.057403}}.
\newline\urlprefix\url{https://link.aps.org/doi/10.1103/PhysRevLett.102.057403}

\bibitem{PhysRevLett.121.060401}
J.~F. Haase, P.~J. Vetter, T.~Unden, A.~Smirne, J.~Rosskopf, B.~Naydenov,
  A.~Stacey, F.~Jelezko, M.~B. Plenio, S.~F. Huelga,
  \href{https://link.aps.org/doi/10.1103/PhysRevLett.121.060401}{Controllable
  non-markovianity for a spin qubit in diamond}, Phys. Rev. Lett. 121 (2018)
  060401.
\newblock \href {https://doi.org/10.1103/PhysRevLett.121.060401}
  {\path{doi:10.1103/PhysRevLett.121.060401}}.
\newline\urlprefix\url{https://link.aps.org/doi/10.1103/PhysRevLett.121.060401}

\bibitem{PhysRevB.81.035205}
M.~Steiner, P.~Neumann, J.~Beck, F.~Jelezko, J.~Wrachtrup,
  \href{https://link.aps.org/doi/10.1103/PhysRevB.81.035205}{Universal
  enhancement of the optical readout fidelity of single electron spins at
  nitrogen-vacancy centers in diamond}, Phys. Rev. B 81 (2010) 035205.
\newblock \href {https://doi.org/10.1103/PhysRevB.81.035205}
  {\path{doi:10.1103/PhysRevB.81.035205}}.
\newline\urlprefix\url{https://link.aps.org/doi/10.1103/PhysRevB.81.035205}

\bibitem{PhysRevB.74.104303}
N.~B. Manson, J.~P. Harrison, M.~J. Sellars,
  \href{https://link.aps.org/doi/10.1103/PhysRevB.74.104303}{Nitrogen-vacancy
  center in diamond: Model of the electronic structure and associated
  dynamics}, Phys. Rev. B 74 (2006) 104303.
\newblock \href {https://doi.org/10.1103/PhysRevB.74.104303}
  {\path{doi:10.1103/PhysRevB.74.104303}}.
\newline\urlprefix\url{https://link.aps.org/doi/10.1103/PhysRevB.74.104303}

\bibitem{vetter2021zero}
P.~J. Vetter, A.~Marshall, G.~T. Genov, T.~F. Weiss, N.~Striegler, E.~F.
  Gro\ss{}mann, S.~Oviedo-Casado, J.~Cerrillo, J.~Prior, P.~Neumann,
  F.~Jelezko,
  \href{https://link.aps.org/doi/10.1103/PhysRevApplied.17.044028}{Zero- and
  low-field sensing with nitrogen-vacancy centers}, Phys. Rev. Applied 17
  (2022) 044028.
\newblock \href {https://doi.org/10.1103/PhysRevApplied.17.044028}
  {\path{doi:10.1103/PhysRevApplied.17.044028}}.
\newline\urlprefix\url{https://link.aps.org/doi/10.1103/PhysRevApplied.17.044028}

\bibitem{frank2017autonomous}
F.~Frank, T.~Unden, J.~Zoller, R.~S. Said, T.~Calarco, S.~Montangero,
  B.~Naydenov, F.~Jelezko, Autonomous calibration of single spin qubit
  operations, npj Quantum Information 3~(1) (2017) 1--5.

\bibitem{haberle2013high}
T.~H{\"a}berle, D.~Schmid-Lorch, K.~Karrai, F.~Reinhard, J.~Wrachtrup,
  High-dynamic-range imaging of nanoscale magnetic fields using optimal control
  of a single qubit, Physical review letters 111~(17) (2013) 170801.

\bibitem{poulsen2022optimal}
A.~F.~L. Poulsen, J.~D. Clement, J.~L. Webb, R.~H. Jensen, L.~Troise,
  K.~Berg-S\o{}rensen, A.~Huck, U.~L. Andersen,
  \href{https://link.aps.org/doi/10.1103/PhysRevB.106.014202}{Optimal control
  of a nitrogen-vacancy spin ensemble in diamond for sensing in the pulsed
  domain}, Phys. Rev. B 106 (2022) 014202.
\newblock \href {https://doi.org/10.1103/PhysRevB.106.014202}
  {\path{doi:10.1103/PhysRevB.106.014202}}.
\newline\urlprefix\url{https://link.aps.org/doi/10.1103/PhysRevB.106.014202}

\bibitem{Li_Ruths_2009}
J.-S. Li, J.~Ruths, D.~Stefanatos, \href{https://doi.org/10.1063/1.3253796}{A
  pseudospectral method for optimal control of open quantum systems}, The
  Journal of Chemical Physics 131~(16) (2009) 164110.
\newblock \href {http://arxiv.org/abs/https://doi.org/10.1063/1.3253796}
  {\path{arXiv:https://doi.org/10.1063/1.3253796}}, \href
  {https://doi.org/10.1063/1.3253796} {\path{doi:10.1063/1.3253796}}.
\newline\urlprefix\url{https://doi.org/10.1063/1.3253796}

\bibitem{Ruths_2011}
J.~Ruths, J.-S. Li, Optimal control of inhomogeneous ensembles, IEEE
  Transactions on Automatic Control 57~(8) (2012) 2021--2032.
\newblock \href {https://doi.org/10.1109/TAC.2012.2195920}
  {\path{doi:10.1109/TAC.2012.2195920}}.

\bibitem{nocedalNumericalOptimization2006}
J.~Nocedal, S.~J. Wright, Numerical Optimization, 2nd Edition, Springer Series
  in Operations Research, {Springer}, {New York}, 2006.

\bibitem{byrdLIMITEDMEMORYALGORITHM}
R.~H. Byrd, P.~Lu, J.~Nocedal, C.~Zhu, \href{https://doi.org/10.1137/0916069}{A
  limited memory algorithm for bound constrained optimization}, SIAM Journal on
  Scientific Computing 16~(5) (1995) 1190--1208.
\newblock \href {http://arxiv.org/abs/https://doi.org/10.1137/0916069}
  {\path{arXiv:https://doi.org/10.1137/0916069}}, \href
  {https://doi.org/10.1137/0916069} {\path{doi:10.1137/0916069}}.
\newline\urlprefix\url{https://doi.org/10.1137/0916069}

\bibitem{moralesRemarkAlgorithm7782011}
J.~L. Morales, J.~Nocedal,
  \href{https://doi.org/10.1145/2049662.2049669}{Remark on “algorithm 778:
  L-bfgs-b: Fortran subroutines for large-scale bound constrained
  optimization”}, ACM Trans. Math. Softw. 38~(1) (dec 2011).
\newblock \href {https://doi.org/10.1145/2049662.2049669}
  {\path{doi:10.1145/2049662.2049669}}.
\newline\urlprefix\url{https://doi.org/10.1145/2049662.2049669}

\bibitem{zhuAlgorithm778LBFGSB1997}
C.~Zhu, R.~H. Byrd, P.~Lu, J.~Nocedal,
  \href{https://doi.org/10.1145/279232.279236}{Algorithm 778: L-bfgs-b: Fortran
  subroutines for large-scale bound-constrained optimization}, ACM Trans. Math.
  Softw. 23~(4) (1997) 550–560.
\newblock \href {https://doi.org/10.1145/279232.279236}
  {\path{doi:10.1145/279232.279236}}.
\newline\urlprefix\url{https://doi.org/10.1145/279232.279236}

\bibitem{virtanenSciPyFundamentalAlgorithms2020}
P.~Virtanen, R.~Gommers, T.~E. Oliphant, M.~Haberland, T.~Reddy, D.~Cournapeau,
  E.~Burovski, P.~Peterson, W.~Weckesser, J.~Bright, S.~J. {van der Walt},
  M.~Brett, J.~Wilson, K.~J. Millman, N.~Mayorov, A.~R.~J. Nelson, E.~Jones,
  R.~Kern, E.~Larson, C.~J. Carey, {\.I}.~Polat, Y.~Feng, E.~W. Moore,
  J.~VanderPlas, D.~Laxalde, J.~Perktold, R.~Cimrman, I.~Henriksen, E.~A.
  Quintero, C.~R. Harris, A.~M. Archibald, A.~H. Ribeiro, F.~Pedregosa, P.~{van
  Mulbregt}, \href{https://doi.org/10.1038/s41592-019-0686-2}{{{SciPy}} 1.0:
  Fundamental algorithms for scientific computing in {{Python}}}, Nat. Methods
  17~(3) (2020) 261--272.
\newblock \href {https://doi.org/10.1038/s41592-019-0686-2}
  {\path{doi:10.1038/s41592-019-0686-2}}.
\newline\urlprefix\url{https://doi.org/10.1038/s41592-019-0686-2}

\bibitem{jax2018github}
J.~Bradbury, R.~Frostig, P.~Hawkins, M.~J. Johnson, C.~Leary, D.~Maclaurin,
  G.~Necula, A.~Paszke, J.~Vander{P}las, S.~Wanderman-{M}ilne, Q.~Zhang,
  \href{http://github.com/google/jax}{{JAX}: composable transformations of
  {P}ython+{N}um{P}y programs} (2018).
\newline\urlprefix\url{http://github.com/google/jax}

\bibitem{defouquieresSecondOrderGradient2011a}
P.~{de Fouquieres}, S.~G. Schirmer, S.~J. Glaser, I.~Kuprov, Second order
  gradient ascent pulse engineering, Journal of Magnetic Resonance 212~(2)
  (2011) 412--417.
\newblock \href {https://doi.org/10.1016/j.jmr.2011.07.023}
  {\path{doi:10.1016/j.jmr.2011.07.023}}.

\bibitem{goodwinAdvancedOptimalControl2017}
D.~L. Goodwin, Advanced {{Optimal Control Methods}} for {{Spin Systems}},
  arXiv:1803.10432 [quant-ph] (2017).
\newblock \href {http://arxiv.org/abs/1803.10432} {\path{arXiv:1803.10432}},
  \href {https://doi.org/10.5258/soton/t0003} {\path{doi:10.5258/soton/t0003}}.

\bibitem{hogbenSpinachSoftwareLibrary2011}
H.~J. Hogben, M.~Krzystyniak, G.~T.~P. Charnock, P.~J. Hore, I.~Kuprov, Spinach
  \textendash{} {{A}} software library for simulation of spin dynamics in large
  spin systems, Journal of Magnetic Resonance 208~(2) (2011) 179--194.
\newblock \href {https://doi.org/10.1016/j.jmr.2010.11.008}
  {\path{doi:10.1016/j.jmr.2010.11.008}}.

\bibitem{floetherRobustQuantumGates2012a}
F.~F. Floether, P.~de~Fouquieres, S.~G. Schirmer,
  \href{https://doi.org/10.1088/1367-2630/14/7/073023}{Robust quantum gates for
  open systems via optimal control: {{Markovian}} versus non-{{Markovian}}
  dynamics}, New J. Phys. 14~(7) (2012) 073023.
\newblock \href {https://doi.org/10.1088/1367-2630/14/7/073023}
  {\path{doi:10.1088/1367-2630/14/7/073023}}.
\newline\urlprefix\url{https://doi.org/10.1088/1367-2630/14/7/073023}

\bibitem{PhysRevA.95.042318}
N.~Leung, M.~Abdelhafez, J.~Koch, D.~Schuster,
  \href{https://link.aps.org/doi/10.1103/PhysRevA.95.042318}{Speedup for
  quantum optimal control from automatic differentiation based on graphics
  processing units}, Phys. Rev. A 95 (2017) 042318.
\newblock \href {https://doi.org/10.1103/PhysRevA.95.042318}
  {\path{doi:10.1103/PhysRevA.95.042318}}.
\newline\urlprefix\url{https://link.aps.org/doi/10.1103/PhysRevA.95.042318}

\end{thebibliography}


\begin{thebibliography}{0}
\bibitem[1]{NumPy} T. E. Oliphant, Comput. Sci. Eng. 9, 10 (2007). \url{http://www.scipy.org/}
\bibitem[2]{JAX} J. Bradbury et al., \textit{{JAX}: composable transformations of {P}ython+{N}um{P}y programs}, (2018), \url{http://github.com/google/jax}
\bibitem[3]{Training_Cat} S. Glaser, U. Boscain, T. Calarco, et al., Eur. Phys. J. D 69, 279 (2015)
\bibitem[4]{Koch} C. P. Koch, U. Boscain, T. Calarco, et al., EPJ Quantum Technol. (2022) 9: 19
\bibitem[5]{dCRAB} N. Rach, M. M. M{\"u}ller, T. Calarco and S. Montangero, Phys. Rev. A 92 (2015), 6
\bibitem[6]{GRAPE} N. Khaneja et al., J. Magn. Reson. 92 (2015), 6, p. 296-305
\bibitem[7]{AD} N. Leung et al., Phys. Rev. A 95 (2017), 4
\bibitem[8]{Qudi} J. M. Binder et al., SoftwareX 6 (2017), p. 85-90
\end{thebibliography}


\section*{Required Metadata}
\label{sec:metadate}
Github repository: \\
\url{https://github.com/Quantum-OCS/QuOCS}

\end{document}